\documentclass[%
 reprint,
 amsmath,amssymb,
 aps,prx,
 twocolumn,
 showpacs,
 superscriptaddress,
 floatfix,
 showkeys,
 reprint,
 longbibliography,]
 {revtex4-2}

\usepackage{graphicx}
\usepackage{dcolumn}
\usepackage{bm}
\usepackage{amsmath,amssymb}
\usepackage{hyperref}
\usepackage{natbib}
\usepackage{xcolor}
\usepackage[normalem]{ulem}
\usepackage{booktabs}
\usepackage{tabularx}
\usepackage[english]{babel}
\usepackage[utf8]{inputenc}
\usepackage{enumerate}
\usepackage{newunicodechar}
\usepackage{lipsum}

\usepackage[font=small,format=plain,labelfont=bf,
singlelinecheck=false,justification=raggedright,
labelsep=space,figurename=FIG.]{caption} 
\usepackage{ragged2e} 

\begin{document}

\title{Electric-field induced trends of exchange interactions in transition-metal trilayers} 

\author{Moinak Ghosh}
\affiliation{Center for High-Performance Computing, Indian Institute of Science Education and Research Thiruvananthapuram, Thiruvananthapuram, Kerala 695551, India}

\author{Stefan Heinze}
\affiliation{Institute of Theoretical Physics and Astrophysics, Christian-Albrechts-Universit$\ddot{a}$t zu Kiel, Leibnizstrasse 15, 24098 Kiel, Germany}
\affiliation{Kiel Nano, Surface, and Interface Science (KiNSIS), Christian-Albrechts-Universit$\ddot{a}$t zu Kiel, Christian-Albrechts-Platz 4, 24118 Kiel, Germany}

\author{Souvik Paul}
\email{souvikpaul@iisertvm.ac.in}
\affiliation{Center for High-Performance Computing, Indian Institute of Science Education and Research Thiruvananthapuram, Thiruvananthapuram, Kerala 695551, India}
\affiliation{School of Physics, Indian Institute of Science Education and Research Thiruvananthapuram, Thiruvananthapuram, Kerala 695551, India}

\date{\today}

\begin{abstract}
	
Using density functional theory, we have performed a systematic study of the Heisenberg pairwise exchange interaction and the beyond Heisenberg multi-spin higher-order exchange interactions in unsupported transition-metal trilayers in the presence of external electric fields. The systems consist of a hexagonal atomic Fe layer sandwiched between 4$d$ (Ru, Rh, and Pd) and 5$d$ (Ir) transition-metal layers. Both fcc and hcp stackings of the 4$d$ overlayer have been taken into account. To scan a large part of the magnetic phase space, we have calculated the energy dispersion of spin spirals without and with applied electric fields up to $\pm 1.0$ V/{\AA}. We find that the energy dispersion remains qualitatively the same upon applying the electric fields and the magnetic ground state remains unchanged. The exchange constants obtained by fitting the energy dispersions exhibit a linear dependence on the electric field up to values of about $\pm 0.5$ V/{\AA}. The sign of the calculated pairwise and higher-order exchange constants remain unchanged with electric field, however, their values and field induced variation are sensitive to the 4$d$ overlayer. The changes are on the order of a few percent for the nearest-neighbor exchange constant and up to a few ten percent for beyond nearest-neighbor constants. The higher-order exchange constants are calculated based on the total energies of multi-$Q$ states, such as the $uudd$ and the 3$Q$ state. Similar to the pairwise exchange constants, we find a nearly linear field dependence of the higher order constants at small electric fields and variations of up to ten percent. We study the spin-dependent screening of the electric field for the three trilayers based on the spin- and orbital-decomposed
electronic states. The modifications of the pairwise and higher-order exchange interaction constants are related to the electric field induced changes of the spin-dependent Fe local density of states and its variation at the Fermi level. Our work shows the significant effect of an electric field on exchange interactions at transition-metal interfaces and emphasizes its importance for electric-field assisted switching of topological spin structures, such as magnetic skyrmions and antiskyrmions, in ultrathin transition-metal films.

\end{abstract}

\maketitle


\section{Introduction}

Since the discovery of topological nontrivial spin structures, such as magnetic skyrmions~\cite{Bogdanov1994,nagaosa13}, in two-dimensional (2D) chiral magnets, e.g., transition-metal ultrathin films~\cite{Heinze2011,Romming2013,Romming2015,Hagemeister:15.1,kubetzka:17.1,Hanneken:15.1,meyer19}, multilayer heterostructures~\cite{Boulle2016,Moreau-Luchaire2016,Woo2016,Soumya2017,chen2015}, and layered van der Waals materials~\cite{han2019,ding2020,park2021,wu2022,powalla2023}, various applications have been proposed, ranging from next-generation spintronic devices~\cite{stuart2008,tomasello14,zhang2015,kang2016,tomasello2017,luo2018,nagaosa13} to neuromorphic~\cite{grollier2020,song2020} and quantum computing~\cite{psaroudaki2021,psaroudaki2022}. These applications are possible due to their topological properties, which provide enhanced stability against external perturbations, and localized nano-sized spin structures, which enable high-density data storage and energy-efficient manipulation. 

A widely used approach to drive and control magnetic skyrmions is via a spin-polarized current, which operates through two mechanisms: spin-transfer torque~\cite{nagaosa13,junichi13,sampaio13,fert13} and spin-orbit torque~\cite{alex2009,miron2010,miron2011}. However, both mechanisms drive skyrmions not only along the current direction but also in the transverse direction. This transverse motion occurs due to the skyrmion Hall effect~\cite{zang2011,matano2016,jiang2017}, which arises due to the nontrivial topology of these spin structures and leads to the undesirable loss of data in the devices. Furthermore, the electric current generates the Joule heating effect, which acts against the thermal stability of skyrmions. In spite of research on skyrmions over more than a decade, these drawbacks have hindered the practical realization of skyrmion-based devices. The electric-field assisted creation and annihilation of skyrmions are elegant solutions to the above problems~\cite{hsu17,schott17,goerzen2022,paul2022}. In this process, an electric field is applied locally to switch skyrmions rather than transport them, leading to a more energy-efficient route for manipulation.

Despite such promise, studies on the electric-field induced creation and annihilation of skyrmions are relatively limited. The experimental studies have focused on the switching of skyrmions and skyrmionic bubbles in transition-metal multilayers~\cite{hsu17,schott17,ma18,titiksha18} and in multiferroic heterostructures~\cite{Wang2020,Ba2021}. Recently, the electric-field induced switching of skyrmion helicity in a multiferroic heterostructure~\cite{dai2023} and in a van der Waals heterostructure~\cite{Han2025} has been reported. The electric field modifies the skyrmion properties via tuning the magnetic interactions, including the Heisenberg pairwise exchange interaction, Dzyaloshinskii-Moriya interaction (DMI), magnetocrystalline anisotropy energy (MAE), and dipole-dipole interaction, as well as the beyond Heisenberg multi-spin exchange interactions~\cite{paulhoi,paul2022}. Theoretical studies initially considered the effect of an electric field on these interactions individually~\cite{upadhyaya15,fook16,nakatani16,oba15,yang2018,titiksha18,desplat2021}, and recently, investigations including their combined effects have been performed~\cite{li2021,paul2022}. Moreover, a recent theoretical study, based on an atomistic spin model, has further demonstrated that pairwise exchange interactions can play an essential role in this switching process~\cite{goerzen2022}. Nevertheless, a systematic and comprehensive study across various systems, as required for practical applications, is still missing.

\begin{figure*}[!htbp]
	\includegraphics[scale=1.0]{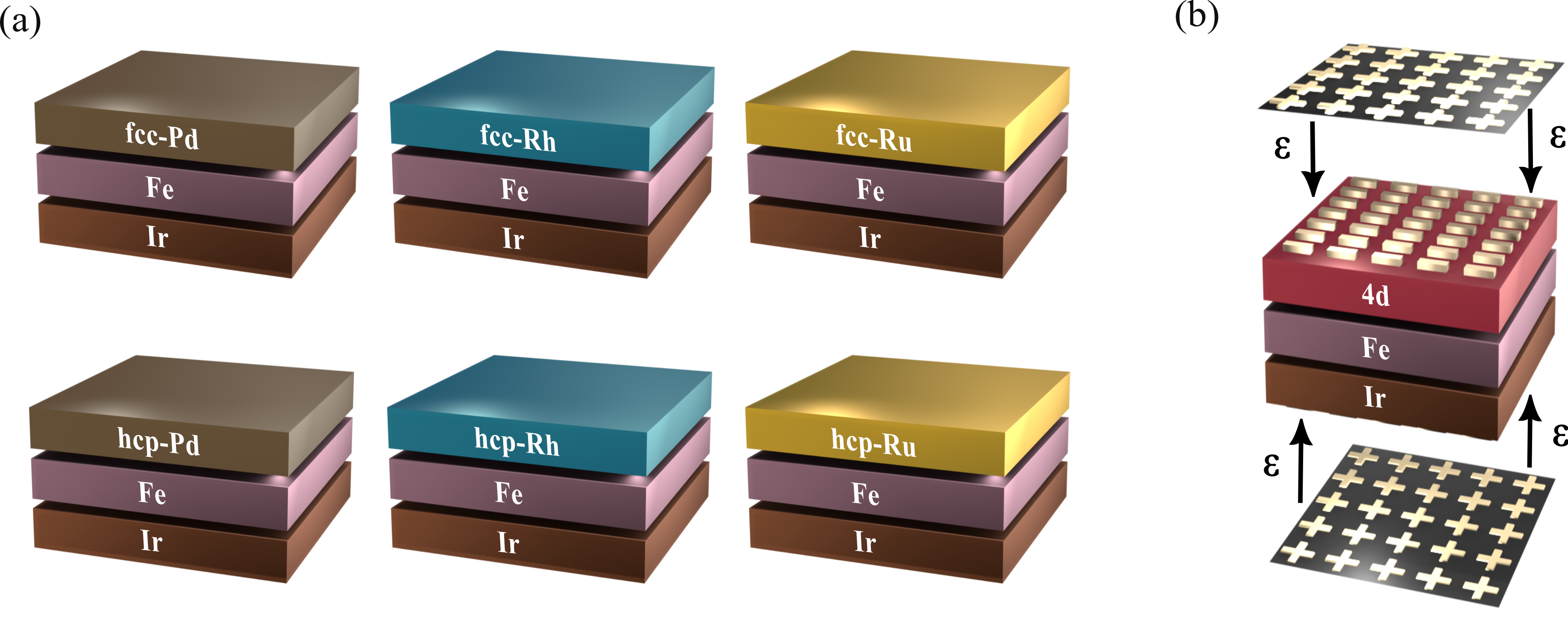}
	\centering
	\caption{\justifying Unsupported 4$d$/Fe/Ir trilayers under an electric field. (a) Top row displays three trilayers: Pd/Fe/Ir, Rh/Fe/Ir, and Ru/Fe/Ir with fcc stacking of the overlayer, while the bottom row displays the trilayers with hcp stacking of the overlayer. (b) Illustration of a symmetric uniform electric field, perpendicular to the 4$d$/Fe/Ir trilayer. The symmetric electric field is created by placing a charged sheet at 5.3 \AA~above the 4$d$ overlayer and below the Ir bottom layer. The directions of field lines for the top and bottom negative electric fields are shown.}
	\label{fig:f1}
\end{figure*}

Here, based on density functional theory (DFT), we systematically analyze the trends in magnetic interactions, such as the Heisenberg pairwise exchange interaction and higher-order exchange interactions (HOI), for freestanding Pd/Fe/Ir and Rh/Fe/Ir trilayers under applied electric fields (Fig.~\ref{fig:f1}(a)). Our choice of trilayers is motivated by the experimental observation of skyrmions and complex spin structures in the corresponding ultrathin films, i.e., Pd/Fe/Ir(111) \cite{Romming2013,Romming2015,Muckel2021} and Rh/Fe/Ir(111) \cite{romming18}. Both the fcc and hcp stackings of the $4d$ transition-metal overlayer are considered, as they have been reported in experiments and because the stacking order can influence the magnetic interactions and ground state of the ultrathin film systems~\cite{dupe2014,malottki2017a,romming18}. Additionally, to examine the role of band filling, the electric-field induced effects on the magnetic properties of Ru/Fe/Ir trilayers, with fcc and hcp stackings of the overlayer, are also investigated. The unsupported trilayers serve as simplified model systems of the corresponding films, in which the general trends can be examined more easily~\cite{gutzeit2021}. It is important to note that the accurate calculation of DMI and MAE depends on the substrate layers, and therefore these interactions are excluded from the present study.

For all trilayers, we perform total energy DFT calculations, without and with electric fields, of the spin spiral (1Q) and multi-$Q$ states, obtained from a superposition of symmetry equivalent spin spirals. The energy dispersion calculated under electric fields is qualitatively the same as that of the zero-field dispersion. For all trilayers, the energy dispersion of spin spirals and the total energies of multi-$Q$ states vary in a similar way under electric fields. However, the magnetic ground state of all trilayers remains unaltered under electric fields up to $\pm 1.0$~V/{\AA}.

From mapping the total DFT energies onto a spin model containing the Heisenberg and higher-order exchange terms, we obtain the corresponding interaction constants. The obtained interaction constants exhibit a linear dependence on the electric field up to values of about $\pm 0.5$~V/{\AA} for all trilayers and deviations in some cases at larger field values. Moreover, the variation of these constants with electric fields follows largely similar trends for Pd/Fe/Ir and Rh/Fe/Ir trilayers with fcc and hcp stackings and deviates for Ru/Fe/Ir trilayers. We explain the field-induced change in the spin-dependent charge densities based on changes in the spin- and orbital-resolved local density of states and relate these to the modifications of the exchange interaction constants.

Our study suggests a significant effect of exchange interactions on the electric-field induced manipulation and switching of topological spin textures, such as magnetic skyrmions and antiskyrmions. Since it has been previously shown that exchange frustration~\cite{malottki2017a} and higher-order exchange~\cite{paulhoi} can have a large effect on skyrmion and antiskyrmion stability, the electric-field induced modifications of these terms can play a considerable role in the switching process. We anticipate that our study will motivate experimentalists to explore the electric-field switching of skyrmions and antiskyrmions in ultrathin transition-metal films at surfaces for which our trilayers represent model systems.

This paper is structured in the following way. Sec.~\ref{sec:compdet} contains the details of the computational method used to calculate the total DFT energies of collinear and noncollinear magnetic states with external electric fields and how the magnetic interactions were obtained from these electronic structure calculations. In the main part of the paper, i.e., Sec.~\ref{sec:resdiss}, we present and discuss the results of different trilayer systems. Finally, we summarize our main results in Sec.~\ref{sec:conc}.                      

\section{\label{sec:compdet} Computational details}

The electronic structure and magnetic properties of the unsupported trilayers were calculated using the full-potential linearized augmented-plane-wave (FLAPW) based \textsc{fleur} code~\cite{fleur}. The code allows us to calculate the total energies of collinear and noncollinear magnetic states, e.g., spin spiral states, of low-dimensional materials, such as transition-metal trilayers in film geometry and with an external electric field~\cite{kurz2004,heide2009,Zimmermann2014,paul2022}. 

The theoretical in-plane lattice constant of Ir, i.e., 2.70~\AA, obtained via local density approximation (LDA)~\cite{dupe2014}, was chosen for all unsupported trilayers studied in this paper following previous work~\cite{dupe2014,gutzeit2021}. Since our goal is to find the trends among the trilayers by only changing the overlayer similar to the study in Ref.~\cite{gutzeit2021}, we kept the interlayer distances fixed to those of fcc and hcp stacked Rh/Fe/Ir(111) films~\cite{romming18}. 

The LDA, as parameterized by Vosko, Wilk, and Nusair, was used for the exchange-correlation functional~\cite{vwn}. We kept the computational parameters of this study same as those in Ref.~\cite{gutzeit2021}, in which these trilayers were investigated without an electric field. The energy cutoff of the plane waves was set to $k_{\rm max}=4.1$~a.u.$^{-1}$ and 1936 $k$-points were used in the irreducible two-dimensional Brillouin zone (2DBZ) to achieve good convergence. The muffin-tin radii of the 4$d$ and 5$d$ elements were fixed to 2.31 a.u., while that for Fe was chosen as 2.23 a.u.

An efficient approach to search for a noncollinear magnetic ground state is to calculate the energy dispersion of spin spirals, which are the general solutions of the Heisenberg model on a periodic 
lattice~\cite{kurz2004}. Consequently, we calculated the total energy of homogeneous flat spin spirals as a function of the wave vector $\textbf{q}$ along the two high-symmetry directions $\overline{\Gamma \mathrm{KM}}$ and $\overline{\Gamma \mathrm{M}}$ of the 2DBZ. To perform these calculations within the chemical unit cell, avoiding the supercell approach, we used the generalized Bloch theorem, as implemented in the \textsc{fleur} code~\cite{kurz2004}.

We mapped the total energies of spin spirals onto the Heisenberg Hamiltonian to calculate the pairwise exchange constants. This spin Hamiltonian is expressed as
\begin{align} \label{eq1}
	\mathcal{H}_{\rm ex} =- \sum_{ij} J_{ij} \ (\textbf{m}_{i}\cdot\textbf{m}_{j})
\end{align}
where $J_{ij}$ are the pairwise exchange interaction constants, and $\textbf{m}_{i}$ ($\textbf{m}_{j}$) denotes the unit vector along the magnetic moment direction at site $i$ ($j$).

The fourth-order perturbative expansion of the hopping parameter over the Coulomb term in the Hubbard model~\cite{hoffmann20} results in three HOI terms: a two-site four spin interaction, widely known as the biquadratic interaction, a three-site four spin interaction~\cite{hoffmann20}, and a four-site four spin interaction~\cite{macD88,takahashi77}. Following previous studies~\cite{paulhoi,gutzeit2021}, here, we considered these interactions within the approximation of minimal distances between the spins, since they arise from the fourth-order perturbation theory. 

The three terms are of the following form
\begin{gather}
	\mathcal{H}_{\mathrm{biquad}} = - B_1 \sum_{\langle ij \rangle} (\textbf{m}_{i} \cdot \textbf{m}_{j})^2 \\
	\mathcal{H}_{\mathrm{3-site}} = - 2 Y_1 \sum_{\langle ijk \rangle} (\textbf{m}_{i} \cdot \textbf{m}_{j}) (\textbf{m}_{j} \cdot \textbf{m}_{k}) \\
	\mathcal{H}_{\mathrm{4-site}} = - K_1 \sum_{\langle ijkl \rangle} [(\textbf{m}_{i} \cdot \textbf{m}_{j}) (\textbf{m}_{k} \cdot \textbf{m}_{l}) \nonumber \\
	+(\textbf{m}_{i} \cdot \textbf{m}_{l}) (\textbf{m}_{j} \cdot \textbf{m}_{k})-(\textbf{m}_{i} \cdot \textbf{m}_{k}) (\textbf{m}_{j} \cdot \textbf{m}_{l})]
\end{gather}
where $B_1$, $Y_1$, and $K_1$ are the corresponding biquadratic, three-site four spin, and four-site four spin interaction constants, respectively. We denote the minimal-distance approximation by $\langle...\rangle$. The three-site four spin and four-site four spin interactions are abbreviated above as 3-site and 4-site, respectively.  

We evaluated these three HOI constants from three multi-Q states: two 2D collinear up-up-down-down ($uudd$) or double row-wise antiferromagnetic (AFM) states~\cite{hardrat2009,Kronelein2018,romming18} and one 3D noncollinear 3$Q$ state~\cite{pkurz,spethmann2020,Nickel2023b}. The $uudd$ states are constructed from the superposition of spin spirals corresponding to the symmetry equivalent $q$ vectors along the $\overline{\Gamma \mathrm{KM}}$ and $\overline{\Gamma \mathrm{M}}$ directions of 2DBZ, while the 3$Q$ state is constructed from the superposition of three spin spirals corresponding to 
the $\overline{\mathrm{M}}$ points of the 2DBZ. The HOI constants are related to the total energy difference between the multi-$Q$ and single-$Q$ (spin spiral) states as follows 
\begin{gather} 
	B_{1}=\frac{3}{32} \Delta E_{\overline{\mathrm{M}}}^{3Q} - \frac{1}{8} \Delta E_{\overline{\mathrm{M}}/2}^{uudd} \label{eq6} \\
	Y_{1}=\frac{1}{8}(\Delta E_{3\overline{\mathrm{K}}/4}^{uudd} - \Delta E_{\overline{\mathrm{M}}/2}^{uudd}) \label{eq7} \\
	K_{1}=\frac{3}{64} \Delta E_{\overline{\mathrm{M}}}^{3Q} + \frac{1}{16} \Delta E_{3{\overline{\mathrm{K}}}/4}^{uudd} \label{eq8}
\end{gather}
where the energy differences are given by
\begin{gather}
	\Delta E_{3\overline{\mathrm{K}}/4}^{uudd}= E_{3{\overline{\mathrm{K}}}/4}^{uudd} - E_{3{\overline{\mathrm{K}}}/4}^{SS} \label{eq9} \\
	\Delta E_{\overline{\mathrm{M}}/2}^{uudd}= E_{\overline{\mathrm{M}}/2}^{uudd} - E_{\overline{\mathrm{M}}/2}^{SS}  \label{eq10} \\
	\Delta E_{\overline{\mathrm{M}}}^{3Q}= E_{\overline{\mathrm{M}}}^{3Q} - E_{\overline{\mathrm{M}}}^{SS}	 \label{eq11}
\end{gather}
where the energy difference between the $uudd$ state and spin spiral state at $3\overline{\mathrm{K}}/4$ along $\overline{\Gamma \mathrm{KM}}$ is denoted as $\Delta E_{3\overline{\mathrm{K}}/4}^{uudd}$, the energy difference between the $uudd$ state and spin spiral state at $\overline{\mathrm{M}}/2$ along $\overline{\Gamma \mathrm{M}}$ is denoted as $\Delta E_{\overline{\mathrm{M}}/2}^{uudd}$, and the energy difference between the 3$Q$ state and the spin spiral state at the $\overline{\mathrm{M}}$ point is denoted as $\Delta E_{\overline{\mathrm{M}}}^{3Q}$.

To maintain the same level of accuracy, the total energies of the multi-$Q$ states were evaluated with the same energy cutoff and muffin-tin radii used for the single-$Q$ states, but the $k$-points were scaled according to the size of the multi-$Q$ state. Similar to Ref.~\cite{gutzeit2021}, we used 168 $k$-points in the irreducible 2DBZ for the $uudd$ state along $\overline{\Gamma \mathrm{M}}$, 336 $k$-points for the $uudd$ state along $\overline{\Gamma \mathrm{KM}}$, and  484 $k$-points for the 3$Q$ state at the $\overline{\mathrm{M}}$ point in our calculations.  

The computational setup used to calculate the energy dispersions of spin spirals and energies of the multi-$Q$ states under an electric field is shown in Fig.~\ref{fig:f1}(b). The symmetric electric field is generated by placing a charged sheet at a distance of 5.3~\AA~on either side of the 
trilayer~\cite{weinert2009,paul2022}. To maintain charge neutrality of the whole setup, an equal and opposite amount of charge is added to or subtracted from the trilayer when the sheets are negatively or positively charged, respectively. The electric field direction for positively charged sheets is shown in Fig.~\ref{fig:f1}(b), which we define as the negative electric field ($\mathcal{E}<0$). The field direction reverses for negatively charged sheets, and consequently, we refer to this as the positive electric field ($\mathcal{E}>0$). The electric field generated in this way is uniform and perpendicular to the trilayer plane.

\begin{figure*}[!htbp]
	\includegraphics[scale=1.0]{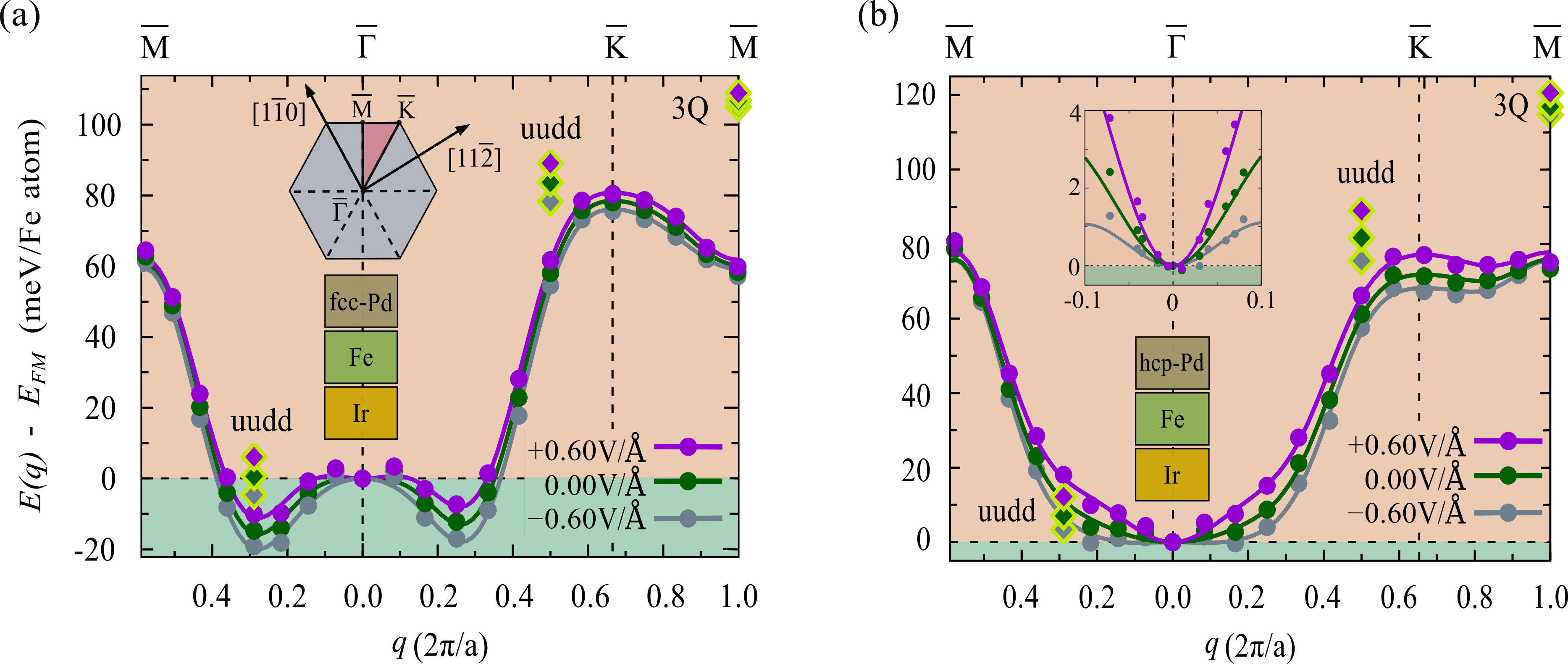}
	\centering
	\caption{\justifying Variation of spin spiral energy dispersion and multi-Q states of Pd/Fe/Ir trilayers with electric fields. Energy dispersion $E$(\textbf{q}) of homogeneous flat spin spirals for (a) the fcc-Pd/Fe/Ir and (b) the hcp-Pd/Fe/Ir trilayers along the two high-symmetry directions $\overline{\Gamma \mathrm{KM}}$ and $\overline{\Gamma \mathrm{M}}$ of the 2DBZ at electric fields of $\mathcal{E}$= 0.6~V/\AA~(violet), 0.0~V/\AA~(green), and $-$0.6~V/\AA~(grey). The filled circles represent DFT data and solid lines are fits to the Heisenberg model (Eq.~(1)). Filled diamonds outlined by solid yellow borders represent the DFT energies of the multi-$Q$ ($uudd$ and 3$Q$) states at the $q$ points of the corresponding single-$Q$ states. The energies of the multi-$Q$ states are also shown for $\mathcal{E}$= 0.0, $\pm 0.6$~V/\AA~(same color scheme as of the single-$Q$ states). Inset of (a) shows a sketch of the 2DBZ with two high-symmetry directions $\overline{\Gamma \mathrm{KM}}$ and $\overline{\Gamma \mathrm{M}}$ and the inset of (b) shows $E$(\textbf{q}) of hcp-Pd/Fe/Ir close to the $\overline{\Gamma}$ point (($\mid$$\textbf{q}$$\mid \leq 0.1\times\frac{2\pi}{a}$)) at $\mathcal{E}$= 0.0, $\pm 0.6$~V/\AA.
    } 
	\label{fig:f2}
\end{figure*}

\section{\label{sec:resdiss} Results and discussion}

Here, we analyze the electric-field induced modifications in the spin spiral energy dispersion $E$(\textbf{q}) of the three unsupported 4$d$/Fe/Ir trilayers, i.e., Pd/Fe/Ir, Rh/Fe/Ir, and Ru/Fe/Ir, with fcc and hcp stacking of the 4$d$ overlayer. We compute the Heisenberg pairwise exchange constants together with the three HOI constants at various electric field values for each trilayer and compare the field-induced trends within the same trilayer type, as well as across the series. To gain more insight into the trends, we further compute the relative change in the magnetic interaction constants with electric fields. Finally, we explain the spin-dependent screening of the electric field via analyzing the field-induced changes in the spin-resolved charge density and relate the charge density difference to the field-induced modifications of the spin-resolved local density of states (LDOS).

\subsection{Pd/Fe/Ir trilayers}

We begin with the DFT energy dispersion $E$(\textbf{q}) of flat spin spirals for the fcc-Pd/Fe/Ir trilayer along the $\overline{\Gamma \mathrm{KM}}$ and $\overline{\Gamma \mathrm{M}}$ directions of the 2DBZ (Fig.~\ref{fig:f2}(a)). The high-symmetry points of the 2DBZ correspond to special magnetic states (inset of Fig.~\ref{fig:f2}(a)). The $\overline{\Gamma}$ point, i.e., $q=0$, represents the ferromagnetic (FM) state, the $\overline{\mathrm{K}}$ point corresponds to the N\'eel state, in which the three adjacent magnetic moments exhibit canting angles of 120$^\circ$, and the  $\overline{\mathrm{M}}$ point resembles the row-wise antiferromagnetic (RW-AFM) state.

Our calculated energy dispersion for fcc-Pd/Fe/Ir at zero field is the same as that obtained in Ref.~\cite{gutzeit2021}. It shows spin spiral energy minima close to the FM state along both directions (Fig.~\ref{fig:f2}(a)). The energy minimum along the $\overline{\Gamma \mathrm{M}}$ direction at $q=0.30 \times \frac{2\pi}{a}$ is lowest between the two directions, and therefore a spin spiral with a period of $\lambda= 0.90 $~nm becomes the magnetic ground state of the trilayer. The energy of the spin spiral ground state is nearly 16 meV/Fe atom below the FM state. The dispersion curve rises sharply thereafter with increasing $q$ value along both directions, and reaches a maximum value of nearly 60 meV/Fe atom at the $\overline{\mathrm{M}}$ point along $\overline{\Gamma \mathrm{M}}$, and nearly 80 meV/Fe atom at the $\overline{\mathrm{K}}$ point along $\overline{\Gamma \mathrm{KM}}$.

We notice that the zero field energy dispersion of the corresponding ultrathin film, i.e., fcc-Pd/Fe/Ir(111), is qualitatively similar to that of the trilayer~\cite{dupe2014,malottki2017a,hoffmann2013}. In particular, energy minima are observed in the vicinity of the FM state along both directions of the energy dispersion for fcc-Pd/Fe/Ir(111), indicating a noncollinear magnetic ground state, similar to the trilayer. However, the depths of these minima in the ultrathin film are only about 1 meV/Fe atom, which is one order of magnitude lower than those of the trilayer. Beyond the spin spiral minima, the energy dispersion of the ultrathin film rises sharply along both directions with increasing $q$ value, similar to the trilayer. For the film system, it reaches a maximum value at the $\overline{\mathrm{K}}$ point along $\overline{\Gamma \mathrm{KM}}$ and at the $\overline{\mathrm{M}}$ point along $\overline{\Gamma \mathrm{M}}$ of around 110 meV/Fe atom higher than the FM state.

\begin{table*} [!htbp]
	\centering
	\caption{\justifying Variation of exchange constants for Pd/Fe/Ir trilayers with electric fields. $i$-th nearest-neighbor pairwise exchange constant ($J_{i}$), three HOI constants, i.e., biquadratic ($B_{1}$), three-site four spin ($Y_{1}$), and four-site four spin ($K_{1}$), and magnetic moment of the Fe layer ($\mu_{\mathrm{Fe}}$) in the FM state for unsupported fcc-Pd/Fe/Ir and hcp-Pd/Fe/Ir trilayers at $\mathcal{E}=0, \pm 0.3, \pm 0.6$~V/\AA. The three nearest-neighbor exchange interaction constants are obtained from mapping the total DFT energy of spin spirals (Fig.~\ref{fig:f2}) onto the Heisenberg model (Eq.~(1)) and the HOI constants are evaluated from multi-$Q$ states using Eqs.~(5)--(10). Electric field values are given in V/\AA, magnetic interaction constants in meV, and magnetic moments in $\mu_{\mathrm{B}}$.}
	\label{tab:table1}
	\begin{ruledtabular}
		\begin{tabular}{cccccccccccc}
			system & $\mathcal{E}$ & $J_{1}$ &  $J_{2}$ & $J_{3}$ &  $B_{1}$ & $Y_{1}$ & $K_{1}$ & $\mu_{\mathrm{Fe}}$\\
			\colrule
			& $-$0.6 & \begin{tabular}[c]{@{}c@{}}11.51\end{tabular} & \begin{tabular}[c]{@{}c@{}}$-$5.88\end{tabular} & \begin{tabular}[c]{@{}c@{}}$-$3.53\end{tabular} & \begin{tabular}[c]{@{}c@{}}2.67\end{tabular} & \begin{tabular}[c]{@{}c@{}}1.14\end{tabular} & \begin{tabular}[c]{@{}c@{}}3.70\end{tabular} & \begin{tabular}[c]{@{}c@{}}2.81\end{tabular} \\
			
			& $-$0.3 & \begin{tabular}[c]{@{}c@{}}11.44\end{tabular} & \begin{tabular}[c]{@{}c@{}}$-$5.77\end{tabular} & \begin{tabular}[c]{@{}c@{}}$-$3.37\end{tabular} & \begin{tabular}[c]{@{}c@{}}2.64\end{tabular} & \begin{tabular}[c]{@{}c@{}}1.21\end{tabular} & \begin{tabular}[c]{@{}c@{}}3.79\end{tabular} & \begin{tabular}[c]{@{}c@{}}2.82\end{tabular}\\
			
			fcc-Pd/Fe/Ir &  0.0 & \begin{tabular}[c]{@{}c@{}}11.38\end{tabular} & \begin{tabular}[c]{@{}c@{}}$-$5.66\end{tabular} & \begin{tabular}[c]{@{}c@{}}$-$3.20\end{tabular} & \begin{tabular}[c]{@{}c@{}}2.62\end{tabular} & \begin{tabular}[c]{@{}c@{}}1.28\end{tabular} & \begin{tabular}[c]{@{}c@{}}3.86\end{tabular} & \begin{tabular}[c]{@{}c@{}}2.82\end{tabular}\\
			
			& $+$0.3 & \begin{tabular}[c]{@{}c@{}}11.32\end{tabular} & \begin{tabular}[c]{@{}c@{}}$-$5.54\end{tabular} & \begin{tabular}[c]{@{}c@{}}$-$3.03\end{tabular} & \begin{tabular}[c]{@{}c@{}}2.61\end{tabular} & \begin{tabular}[c]{@{}c@{}}1.35\end{tabular} & \begin{tabular}[c]{@{}c@{}}3.93\end{tabular} & \begin{tabular}[c]{@{}c@{}}2.82\end{tabular}\\
			
			& $+$0.6 & \begin{tabular}[c]{@{}c@{}}11.26\end{tabular} & \begin{tabular}[c]{@{}c@{}}$-$5.41\end{tabular} & \begin{tabular}[c]{@{}c@{}}$-$2.86\end{tabular} & \begin{tabular}[c]{@{}c@{}}2.59\end{tabular} & \begin{tabular}[c]{@{}c@{}}1.41\end{tabular} & \begin{tabular}[c]{@{}c@{}}4.00\end{tabular} & \begin{tabular}[c]{@{}c@{}}2.83\end{tabular}\\
			\colrule
			& $-$0.6 & \begin{tabular}[c]{@{}c@{}}9.92\end{tabular} & \begin{tabular}[c]{@{}c@{}}$-$2.15\end{tabular} & \begin{tabular}[c]{@{}c@{}}$-$2.94\end{tabular} & \begin{tabular}[c]{@{}c@{}}4.29\end{tabular} & \begin{tabular}[c]{@{}c@{}}2.66\end{tabular} & \begin{tabular}[c]{@{}c@{}}3.05\end{tabular} & \begin{tabular}[c]{@{}c@{}}2.76\end{tabular}\\
			
			& $-$0.3 & \begin{tabular}[c]{@{}c@{}}9.87\end{tabular} & \begin{tabular}[c]{@{}c@{}}$-$2.09\end{tabular} & \begin{tabular}[c]{@{}c@{}}$-$2.68\end{tabular} & \begin{tabular}[c]{@{}c@{}}4.46\end{tabular} & \begin{tabular}[c]{@{}c@{}}2.89\end{tabular} & \begin{tabular}[c]{@{}c@{}}3.18\end{tabular} & \begin{tabular}[c]{@{}c@{}}2.77\end{tabular}\\
			
			hcp-Pd/Fe/Ir &  0.0 & \begin{tabular}[c]{@{}c@{}}9.82\end{tabular} & \begin{tabular}[c]{@{}c@{}}$-$2.02\end{tabular} & \begin{tabular}[c]{@{}c@{}}$-$2.42\end{tabular} & \begin{tabular}[c]{@{}c@{}}4.64\end{tabular} & \begin{tabular}[c]{@{}c@{}}3.13\end{tabular} & \begin{tabular}[c]{@{}c@{}}3.31\end{tabular} & \begin{tabular}[c]{@{}c@{}}2.78\end{tabular}\\
			
			& $+$0.3 & \begin{tabular}[c]{@{}c@{}}9.78\end{tabular} & \begin{tabular}[c]{@{}c@{}}$-$1.93\end{tabular} & \begin{tabular}[c]{@{}c@{}}$-$2.17\end{tabular} & \begin{tabular}[c]{@{}c@{}}4.82\end{tabular} & \begin{tabular}[c]{@{}c@{}}3.36\end{tabular} & \begin{tabular}[c]{@{}c@{}}3.43\end{tabular} & \begin{tabular}[c]{@{}c@{}}2.79\end{tabular}\\
			
			& $+$0.6 & \begin{tabular}[c]{@{}c@{}}9.74\end{tabular} & \begin{tabular}[c]{@{}c@{}}$-$1.84\end{tabular} & \begin{tabular}[c]{@{}c@{}}$-$1.91\end{tabular} & \begin{tabular}[c]{@{}c@{}}5.00\end{tabular} & \begin{tabular}[c]{@{}c@{}}3.59\end{tabular} & \begin{tabular}[c]{@{}c@{}}3.54\end{tabular} & \begin{tabular}[c]{@{}c@{}}2.79\end{tabular}\\
			
			\end{tabular}
	\end{ruledtabular}
\end{table*}

The $uudd$ state along $\overline{\Gamma \mathrm{KM}}$ and $\overline{\Gamma \mathrm{M}}$, and the 3$Q$ state at the $\overline{\mathrm{M}}$ point  are energetically higher than the corresponding single-$Q$ (spin spiral) states at zero field (Fig.~\ref{fig:f2}(a)). The energy difference between the RW-AFM and 3$Q$ states is significantly larger than that between the two $uudd$ states and their corresponding single-$Q$ states ($90^\circ$ spin spirals). These trends remain unchanged in the corresponding ultrathin film fcc-Pd/Fe/Ir(111)~\cite{paulhoi}.

The energy dispersion for the fcc-Pd/Fe/Ir trilayer calculated at electric field values of $\mathcal{E}= \pm 0.6$~V/\AA~remains qualitatively similar to that at zero field (Fig.~\ref{fig:f2}(a)). Specifically, the dispersion at the negative electric field ($\mathcal{E}=-0.6$~V/\AA) is energetically lower, whereas the one at the positive field ($\mathcal{E}=0.6$~V/\AA) is energetically higher than the zero field dispersion. The $q$ value corresponding to the spin spiral energy minimum at zero field remains unchanged after the application of electric fields. However, the depth of the minimum varies by nearly 10 meV/Fe atom between $\mathcal{E}= \pm 0.6$~V/\AA.

The effect of the electric field on the energy dispersion of the fcc-Pd/Fe/Ir(111) ultrathin film is similar to that of the trilayer~\cite{goerzen2022,hoffmann2013}. The energy dispersion obtained at finite electric field values remains qualitatively the same as that at zero field. Similar to the trilayer, the energy dispersion calculated at the positive (negative) field remains energetically higher (lower) than the zero field dispersion. The depth of the energy minima varies with electric fields, however, the change is one order of magnitude smaller in the ultrathin film compared to that of the trilayer between $\mathcal{E}= \pm 0.6$~V/\AA.      

The energy of the two $uudd$ states for the fcc-Pd/Fe/Ir trilayer reduces with $\mathcal{E}$ = $-0.6$~V/\AA~and increases with $\mathcal{E}$ = $0.6$~V/\AA~relative to their zero field value (Fig.~\ref{fig:f2}(a)). In contrast, the energy of the 3$Q$ state varies minimally with $\mathcal{E}=\pm 0.6$~V/\AA. Nevertheless, all the multi-$Q$ states remain energetically higher than their corresponding single-$Q$ states at all electric field values.

\begin{figure*}[!htbp]
	\includegraphics[scale=1.0]{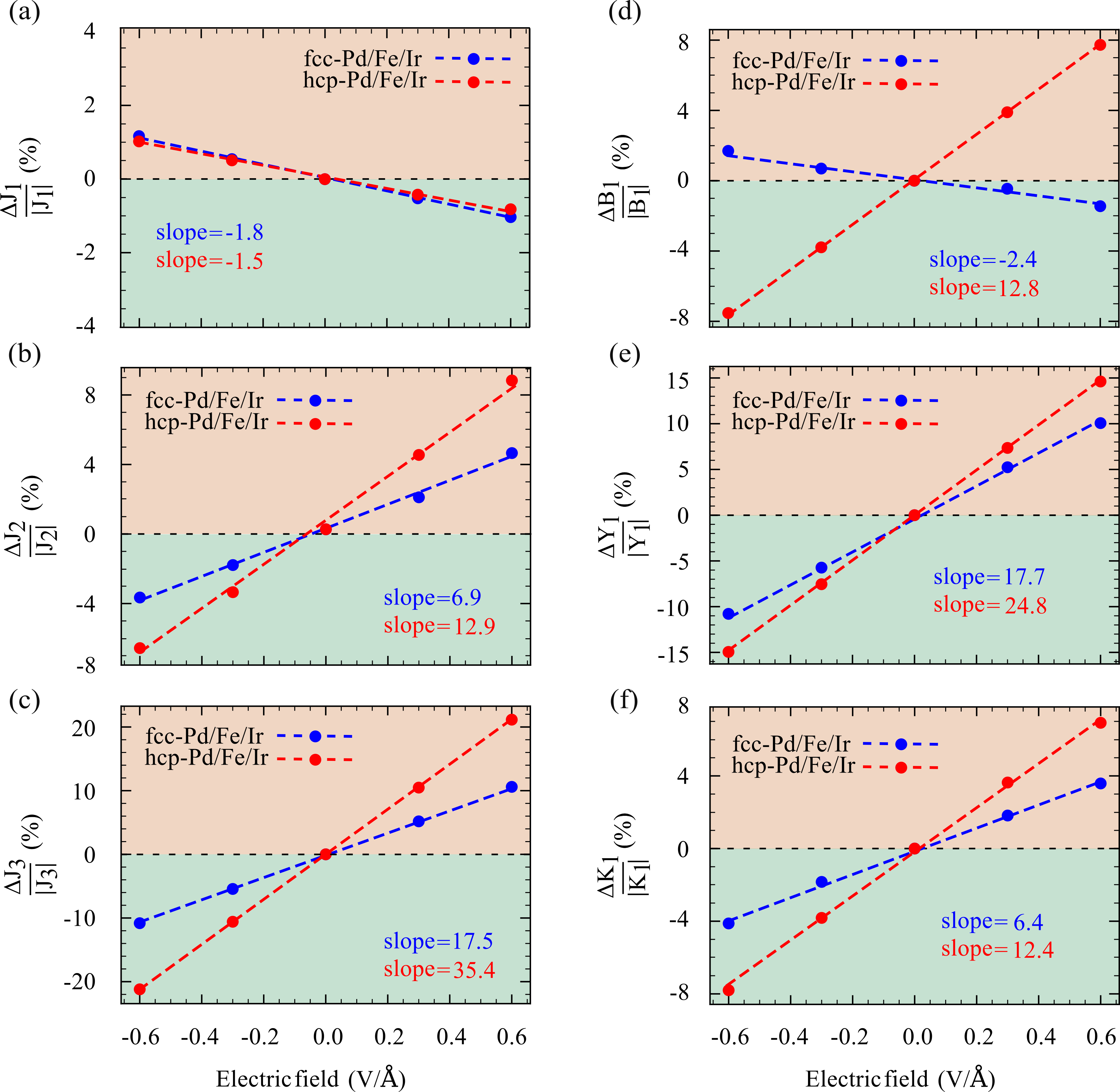}
	\centering
	\caption{\justifying Relative changes of exchange constants with electric fields for Pd/Fe/Ir trilayers. (a) first ($J_1$), (b) second ($J_2$), and (c) third ($J_3$) nearest-neighbor exchange interaction constants, and three HOI constants, i.e., (d) biquadratic ($B_1$), (e) three-site four spin ($Y_1$), and (f) four-site four spin ($K_1$), at two positive ($0.3$~V/\AA~and $0.6$~V/\AA) and two negative ($-0.3$~V/\AA~and $-0.6$~V/\AA) electric fields with reference to the zero electric field for fcc-Pd/Fe/Ir (blue) and hcp-Pd/Fe/Ir (red) trilayers. The solid circles are DFT data and the dashed lines are obtained from a least square fit. The slope of each line is shown and given in \%/(V/\AA).}
	\label{fig:f3}
\end{figure*}

The energy dispersion of Pd/Fe/Ir is modified significantly upon changing the stacking of the Pd overlayer from fcc to hcp (Fig.~\ref{fig:f2}(a) versus Fig.~\ref{fig:f2}(b)). No spin spiral minimum is observed in the energy dispersion of the hcp-Pd/Fe/Ir trilayer at zero field, resulting in an FM ground state (inset of Fig.~\ref{fig:f2}(b)). The dispersion is very flat around the FM state and rises slowly with increasing $q$ value along both directions. It reaches a maximum value at the $\overline{\mathrm{K}}$ point along $\overline{\Gamma \mathrm{KM}}$ and at the BZ boundary along $\overline{\Gamma \mathrm{M}}$ both of nearly 80 meV/Fe atom higher than the FM state.

The energy dispersion of the hcp-Pd/Fe/Ir trilayer at zero field is qualitatively similar to that of the hcp-Pd/Fe/Ir(111) ultrathin film~\cite{dupe2014,malottki2017a}. The ground state of the film is ferromagnetic and the energy dispersion is flat around the FM state. Similar to the trilayer, the maximum of the energy dispersion of the film occurs at the $\overline{\mathrm{M}}$ point along $\overline{\Gamma \mathrm{M}}$. However, in contrast to the trilayer, the maximum energy along $\overline{\Gamma \mathrm{KM}}$ is reached at the $\overline{\mathrm{M}}$ point. Both of these maxima are nearly 110 meV/Fe atom higher than the FM state.

The qualitative nature of the dispersion curve for the hcp-Pd/Fe/Ir trilayer obtained with $\mathcal{E}= \pm 0.6$~V/\AA~remains the same as that obtained at the zero field (Fig.~\ref{fig:f2}(b)). Similar to the fcc stacking, the energy dispersion obtained with negative electric field remains energetically lower and that with positive electric field is energetically higher than the zero field dispersion. The dispersion curves obtained with finite electric fields for hcp-Pd/Fe/Ir(111) ultrathin film exhibit similar trends to those of the trilayer~\cite{goerzen2022,hoffmann2013}.

At zero field, the $uudd$ state of hcp-Pd/Fe/Ir along $\overline{\Gamma \mathrm{KM}}$ and the 3$Q$ state at the $\overline{\mathrm{M}}$ point are energetically higher than their corresponding single-$Q$ states, while the $uudd$ state along $\overline{\Gamma \mathrm{M}}$ is slightly lower in energy than its respective single-$Q$ state (Fig.~\ref{fig:f2}(b)). The total energies of these multi-$Q$ states vary in a similar manner with electric fields as that of the spin spiral dispersion.

To find the trends in the field-induced changes of magnetic interactions, in addition to the above two finite electric field values, we also calculated the energy dispersion of spin spirals and total energies of the multi-$Q$ states at intermediate fields of $\mathcal{E}$ = $\pm 0.3$~V/\AA~for both stackings of the overlayer Pd (not shown). The pairwise exchange constants were evaluated by mapping the total spin spiral energies onto the Heisenberg Hamiltonian (Eq.~(1)) and the HOI constants were obtained using Eqs.~(5)--(10). The first three nearest-neighbor pairwise exchange constants and the three HOI constants at zero and four finite electric field values for the fcc-Pd/Fe/Ir and the hcp-Pd/Fe/Ir trilayers are listed in Table.~\ref{tab:table1}. For completeness, fourth to eighth nearest-neighbor pairwise exchange constants are listed in Table~\ref{apptab:table1} of Appendix~\ref{appsec:beyond_j3}, and the energy difference between the multi-$Q$ and single-$Q$ states used to determine the HOI is provided in Table~\ref{apptab:table2} of Appendix~\ref{appsec:multiq-singleq}. 

The sign of the nearest-neighbor exchange constant ($J_1$) for fcc-Pd/Fe/Ir at zero field is positive, favoring FM coupling, while the second ($J_2$) and third ($J_3$) nearest-neighbor exchange constants possess negative signs, favoring AFM coupling (Table~\ref{tab:table1}). The magnitudes of $J_2$ and $J_3$ are smaller than $J_1$ by factors of nearly 2 and 4, respectively. The existence of competing FM and AFM interactions indicates the presence of exchange frustration, which stabilizes the spin spiral states energetically below the FM state (Fig.~\ref{fig:f2}(a)). 

The signs of these constants remain unchanged in the fcc-Pd/Fe/Ir(111) ultrathin film; however, their values differ modestly from the trilayer. Compared to the trilayer, the value of $J_1$ increases by a factor of about 1.3 in the ultrathin film, while the values of $J_2$ and $J_3$ decrease by factors of nearly 0.4 and 0.8, respectively, in the ultrathin film~\cite{dupe2014,malottki2017a,paulhoi,gutzeit2021}. These changes reflect reduced exchange frustration in the ultrathin film. 

The signs of the three HOI constants --- biquadratic ($B_1$), three-site four spin ($Y_1$), and four-site four spin ($K_1$) --- for fcc-Pd/Fe/Ir are positive at zero field, and their values are comparable with $J_2$ and $J_3$ (Table~\ref{tab:table1}). As observed for the exchange constants, the signs of these three HOI constants remain unchanged in the fcc-Pd/Fe/Ir(111) ultrafilm~\cite{paulhoi,gutzeit2021}, but their values differ moderately between the film and trilayer. The value of $B_1$ in the ultrathin film is higher by a factor of nearly 1.1 compared to the trilayer, whereas the values of $Y_1$ and $K_1$ are lower by a factor of nearly 0.6.

The signs of all interaction constants calculated at zero field remain unchanged under finite electric fields (Table~\ref{tab:table1}). The values of the three exchange constants increase with increasing strength of the negative electric field and decrease with increasing strength of the positive electric field. Among the three HOI constants, $Y_1$ and $K_1$ exhibit an opposite trend, and $B_1$ exhibits the same trend with electric field as that of the exchange constants. Note that the magnetic moment of the Fe layer is about 2.8 $\mu_{\mathrm{B}}$ at zero field and remains almost unaltered under the considered electric field values.

Next, we discuss the effect of electric fields on the exchange and HOI constants of the hcp-Pd/Fe/Ir trilayer (Table~\ref{tab:table1}). The sign of $J_1$ for hcp-Pd/Fe/Ir at zero field is positive, but its magnitude is slightly smaller than that of fcc-Pd/Fe/Ir. The signs of $J_2$ and $J_3$ are negative, but their values are notably smaller than those of fcc-Pd/Fe/Ir, resulting in a weaker exchange frustration. As a result, no spin spiral states are stabilized in hcp-Pd/Fe/Ir, and the ground state is ferromagnetic (Fig.~\ref{fig:f2}(b)). The signs of the three HOI constants of hcp-Pd/Fe/Ir at zero field are positive and their values are higher than $J_2$ and $J_3$, but lower than $J_1$. 

Similar to the fcc-Pd/Fe/Ir trilayer, the signs of these interaction constants at zero field are the same as in the hcp-Pd/Fe/Ir(111) ultrathin film~\cite{malottki2017a,gutzeit2021}, and their values vary up to a modest amount between film and trilayer. The values of $J_1$ and $J_3$ in the ultrathin film increase by factors of nearly 1.4 and 1.2, respectively, relative to those of the trilayer, and the value of $J_2$ decreases by a factor of nearly 0.3. Compared to the trilayer, the value of $B_1$ in the ultrathin film is lower by a factor of nearly 0.3, whereas the values of $Y_1$ and $K_1$ are lower by a factor of nearly 0.7~\cite{malottki2017a,gutzeit2021}.

The signs of the three exchange and three HOI constants for hcp-Pd/Fe/Ir at zero field remain unchanged at finite electric field values. The values of the three exchange interaction constants and the two HOI constants, i.e., $Y_1$ and $K_1$, vary in an analogous way with electric field as those of the fcc-Pd/Fe/Ir trilayer. In contrast, the value of $B_1$ varies with electric field similar to the other two HOI constants, differing from the trend observed in the fcc-Pd/Fe/Ir trilayer. The magnetic moment of the Fe layer at zero field remains very close to that of the fcc-Pd/Fe/Ir trilayer, and its field induced change is negligible for the considered field values. 

Now we turn to the relative change in the three exchange and three HOI constants at finite electric fields with respect to zero field for the fcc-Pd/Fe/Ir and hcp-Pd/Fe/Ir trilayers (Fig.~\ref{fig:f3}). This comparison provides a qualitative assessment of the variation in the interaction constants under electric fields, which is difficult to comprehend from Table~\ref{tab:table1}.

\begin{figure*}[!htbp]
	\includegraphics[scale=1.0]{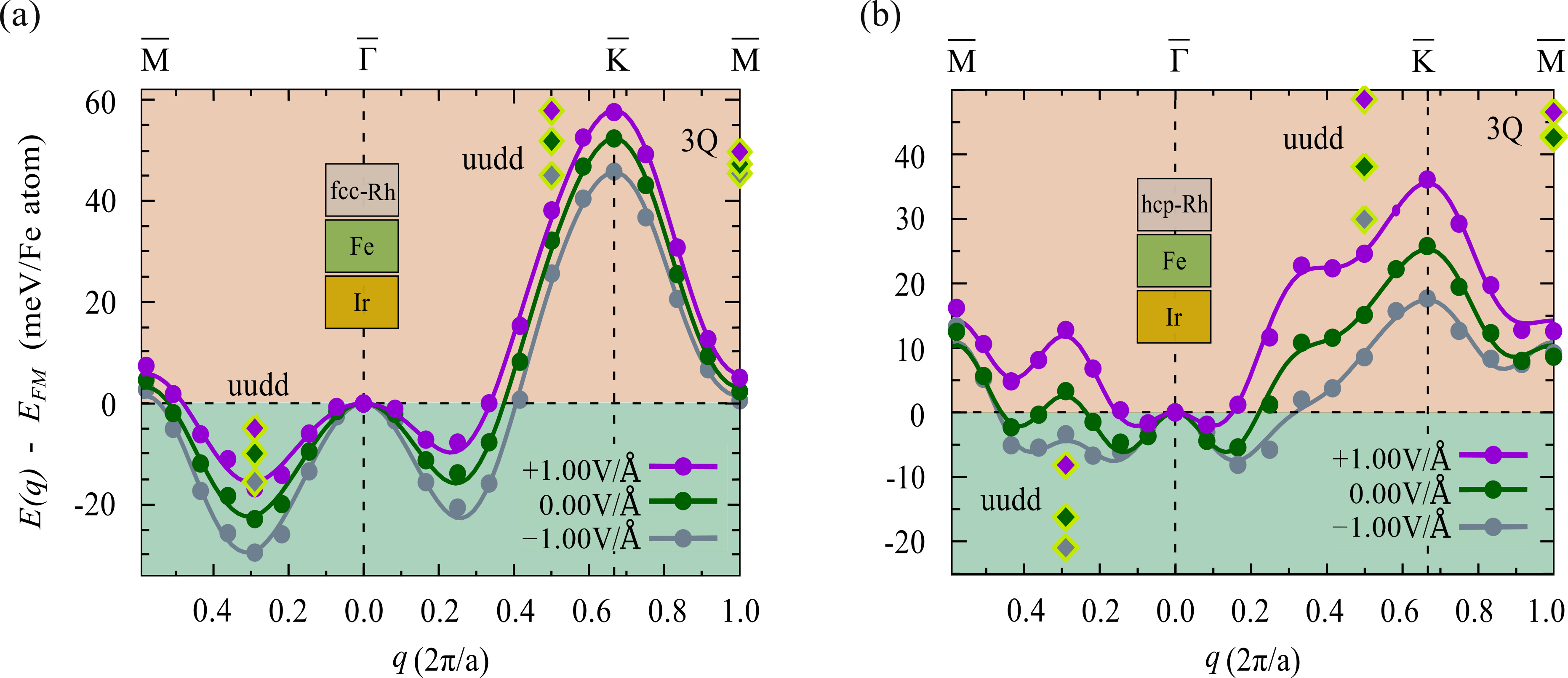}
	\centering
	\caption{\justifying Variation of spin spiral energy dispersion and multi-$Q$ states for Rh/Fe/Ir trilayers with electric fields. Energy dispersion $E$(\textbf{q}) of homogeneous flat spin spiral for (a) the fcc-Rh/Fe/Ir and (b) the hcp-Rh/Fe/Ir trilayer along the two high-symmetry directions $\overline{\Gamma \mathrm{KM}}$ and $\overline{\Gamma \mathrm{M}}$ of the 2DBZ at electric fields of $\mathcal{E}$= 1.0~V/\AA~(violet), 0.0~V/\AA~(green), and $-$1.0~V/\AA~(grey).The filled circles present DFT data and solid lines are fit to the Heisenberg model (Eq.~(1)). Filled diamonds outlined by solid yellow borders represent the DFT energies of the multi-$Q$ ($uudd$ and 3$Q$) states at the $q$ points of the corresponding single-$Q$ states. The energies of the multi-$Q$ states are also shown for $\mathcal{E}$= 0.0, $\pm 1.0$~V/\AA~(same color scheme as of the single-$Q$ states).} 
	\label{fig:f4}
\end{figure*} 

We start with the fcc-Pd/Fe/Ir trilayer (blue symbols and lines in Fig.~\ref{fig:f3}). Overall, the relative changes in all six magnetic interaction constants show a linear dependence on the electric field. The slopes corresponding to the nearest-neighbor exchange constant ($J_1$) and the biquadratic constant ($B_1$) are negative, while the second ($J_2$) and third ($J_3$) nearest-neighbor exchange constants, the three-site four spin ($Y_1$) and four-site four spin ($K_1$) constants exhibit a positive slope, consistent with the above discussion of the data in Table~\ref{tab:table1}. The relative changes in $J_1$ and $B_1$ are small, leading to a slope magnitude on the order of 
2~\%/(V/\AA), those in $J_2$ and $K_1$ are moderate, yielding a slope magnitude on the order of 6~\%/(V/\AA), and those in $J_3$ and $Y_1$ are quite high, exhibiting a slope of one order of magnitude larger.

In general, the magnetic interaction constants of hcp-Pd/Fe/Ir (red symbols and lines in Fig.~\ref{fig:f3}) exhibit a stronger variation with electric fields compared to those of fcc-Pd/Fe/Ir. Specifically, the slope values corresponding to the three HOI constants and $J_2$ and $J_3$ vary between 12 and 35~\%/(V/\AA), whereas $J_1$ exhibits a weak variation, resulting in a small slope value on the order of 2~\%/(V/\AA). Similar to fcc-Pd/Fe/Ir, $J_2$, $J_3$, $Y_1$, and $K_1$ exhibit positive slopes, while $J_1$ displays a negative slope. In contrast to the fcc-Pd/Fe/Ir trilayer, the slope of $B_1$ is positive.

We observed that the energy dispersion of spin spiral and energies of the multi-$Q$ states are modified by an external electric field in a similar manner for both fcc-Pd/Fe/Ir and hcp-Pd/Fe/Ir trilayers. Consequently, the exchange interaction and HOI constants, obtained from the total DFT energies, vary from small to large amounts under electric fields. Interestingly, these interaction constants follow a linear variation with the electric fields up to the considered values of $\mathcal{E}=\pm 0.6$~V/\AA.

\subsection{Rh/Fe/Ir trilayers}

\begin{table*} [!htbp]
	\centering
	\caption{\justifying  Variation of exchange constants for Rh/Fe/Ir trilayers with electric fields. $i$-th nearest-neighbor pairwise exchange constant ($J_{i}$), three HOI constants, i.e., biquadratic ($B_{1}$), three-site four spin ($Y_{1}$), and four-site four spin ($K_{1}$), and magnetic moment of the Fe layer ($\mu_{\mathrm{Fe}}$) in the FM state for unsupported fcc-Rh/Fe/Ir and hcp-Rh/Fe/Ir trilayers at $\mathcal{E}= 0, \pm 0.5, \pm 1.0$~V/\AA. The three nearest-neighbor exchange interaction constants are obtained from mapping the total DFT energy of spin spirals (Fig.~\ref{fig:f4}) onto the Heisenberg model (Eq.~(1)) and the HOI constants are evaluated from multi-$Q$ states using Eqs.~(5)--(10). Electric field values are given in V/\AA, magnetic interaction constants in meV, and magnetic moments in $\mu_{\mathrm{B}}$.}
	\label{tab:table2}
	\begin{ruledtabular}
		\begin{tabular}{ccccccccccccccccc}
			system & $\mathcal{E}$ & $J_{1}$ & $J_{2}$ & $J_{3}$ & $B_{1}$ & $Y_{1}$ & $K_{1}$ & $\mu_{\mathrm{Fe}}$\\
			\colrule
			& $-$1.0 & \begin{tabular}[c]{@{}c@{}}5.69\end{tabular} & \begin{tabular}[c]{@{}c@{}}$-$5.72\end{tabular} & \begin{tabular}[c]{@{}c@{}}$-$0.09\end{tabular} & \begin{tabular}[c]{@{}c@{}}2.46\end{tabular} & \begin{tabular}[c]{@{}c@{}}0.67\end{tabular} & \begin{tabular}[c]{@{}c@{}}3.31\end{tabular} & \begin{tabular}[c]{@{}c@{}}2.44\end{tabular}\\
			
			& $-$0.5 & \begin{tabular}[c]{@{}c@{}}5.71\end{tabular} & \begin{tabular}[c]{@{}c@{}}$-$5.62\end{tabular} & \begin{tabular}[c]{@{}c@{}}0.25\end{tabular} & \begin{tabular}[c]{@{}c@{}}2.52\end{tabular} & \begin{tabular}[c]{@{}c@{}}0.76\end{tabular} & \begin{tabular}[c]{@{}c@{}}3.33\end{tabular}  & \begin{tabular}[c]{@{}c@{}}2.44\end{tabular}\\
			
			fcc-Rh/Fe/Ir & 0.0 & \begin{tabular}[c]{@{}c@{}}5.76\end{tabular} & \begin{tabular}[c]{@{}c@{}}$-$5.50\end{tabular} & \begin{tabular}[c]{@{}c@{}}0.55\end{tabular} & \begin{tabular}[c]{@{}c@{}}2.59\end{tabular} & \begin{tabular}[c]{@{}c@{}}0.84\end{tabular} & \begin{tabular}[c]{@{}c@{}}3.33\end{tabular} & \begin{tabular}[c]{@{}c@{}}2.43\end{tabular}\\
			
			& $+$0.5 & \begin{tabular}[c]{@{}c@{}}5.83\end{tabular} & \begin{tabular}[c]{@{}c@{}}$-$5.36\end{tabular} & \begin{tabular}[c]{@{}c@{}}0.82\end{tabular} & \begin{tabular}[c]{@{}c@{}}2.64\end{tabular} & \begin{tabular}[c]{@{}c@{}}0.91\end{tabular} & \begin{tabular}[c]{@{}c@{}}3.33\end{tabular} & \begin{tabular}[c]{@{}c@{}}2.43\end{tabular}\\
			
			& $+$1.0 & \begin{tabular}[c]{@{}c@{}}5.90\end{tabular} & \begin{tabular}[c]{@{}c@{}}$-$5.21\end{tabular} & \begin{tabular}[c]{@{}c@{}}1.05\end{tabular} & \begin{tabular}[c]{@{}c@{}}2.69\end{tabular} & \begin{tabular}[c]{@{}c@{}}0.98\end{tabular} & \begin{tabular}[c]{@{}c@{}}3.32\end{tabular} & \begin{tabular}[c]{@{}c@{}}2.43\end{tabular}\\
			\colrule
			& $-$1.0 & \begin{tabular}[c]{@{}c@{}}2.13\end{tabular} & \begin{tabular}[c]{@{}c@{}}$-$1.24\end{tabular} & \begin{tabular}[c]{@{}c@{}}0.01\end{tabular} & \begin{tabular}[c]{@{}c@{}}5.37\end{tabular} & \begin{tabular}[c]{@{}c@{}}4.88\end{tabular} & \begin{tabular}[c]{@{}c@{}}2.91\end{tabular} & \begin{tabular}[c]{@{}c@{}}2.43\end{tabular}\\
			
			& $-$0.5 & \begin{tabular}[c]{@{}c@{}}2.16\end{tabular} & \begin{tabular}[c]{@{}c@{}}$-$1.20\end{tabular} & \begin{tabular}[c]{@{}c@{}}0.56\end{tabular} & \begin{tabular}[c]{@{}c@{}}5.51\end{tabular} & \begin{tabular}[c]{@{}c@{}}5.11\end{tabular} & \begin{tabular}[c]{@{}c@{}}2.97\end{tabular} & \begin{tabular}[c]{@{}c@{}}2.44\end{tabular}\\
			
			hcp-Rh/Fe/Ir & 0.0 & \begin{tabular}[c]{@{}c@{}}2.20\end{tabular} & \begin{tabular}[c]{@{}c@{}}$-$1.14\end{tabular} & \begin{tabular}[c]{@{}c@{}}1.07\end{tabular} & \begin{tabular}[c]{@{}c@{}}5.63\end{tabular} & \begin{tabular}[c]{@{}c@{}}5.31\end{tabular} & \begin{tabular}[c]{@{}c@{}}3.02\end{tabular} & \begin{tabular}[c]{@{}c@{}}2.46\end{tabular}\\
			
			& $+$0.5 & \begin{tabular}[c]{@{}c@{}}2.26\end{tabular} & \begin{tabular}[c]{@{}c@{}}$-$1.05\end{tabular} & \begin{tabular}[c]{@{}c@{}}1.53\end{tabular} & \begin{tabular}[c]{@{}c@{}}5.73\end{tabular} & \begin{tabular}[c]{@{}c@{}}5.48\end{tabular} & \begin{tabular}[c]{@{}c@{}}3.05\end{tabular} & \begin{tabular}[c]{@{}c@{}}2.48\end{tabular}\\
			
			& $+$1.0 & \begin{tabular}[c]{@{}c@{}}2.34\end{tabular} & \begin{tabular}[c]{@{}c@{}}$-$0.92\end{tabular} & \begin{tabular}[c]{@{}c@{}}1.94\end{tabular} & \begin{tabular}[c]{@{}c@{}}5.80\end{tabular} & \begin{tabular}[c]{@{}c@{}}5.61\end{tabular} & \begin{tabular}[c]{@{}c@{}}3.07\end{tabular} & \begin{tabular}[c]{@{}c@{}}2.50\end{tabular}\\
			
			\end{tabular}
	\end{ruledtabular}
\end{table*}    

Next, we replace the Pd overlayer with its preceding 4$d$ element in the periodic table, and study Rh/Fe/Ir trilayer. Both the fcc and hcp stackings of the Rh overlayer are considered, since they have been observed experimentally in film geometry~\cite{romming18}.

The energy dispersions of spin spirals for the fcc-Rh/Fe/Ir and hcp-Rh/Fe/Ir trilayers (Figs.~\ref{fig:f4}(a) and~\ref{fig:f4}(b) ) are qualitatively similar to that of fcc-Pd/Fe/Ir: spin spiral minima are observed along both high-symmetry directions of the 2DBZ close to the FM state ($\overline{\Gamma}$ point) and the energy maximum occurs at the $\overline{\mathrm{K}}$ point, representing the N\'eel state. 

For fcc-Rh/Fe/Ir, the energy minimum at zero field occurs along $\overline{\Gamma \mathrm{M}}$ at $q=0.30 \times \frac{2\pi}{a}$, which corresponds to a spin spiral period of $\lambda= 0.90$~nm. This spin spiral minimum is nearly 22 meV/Fe atom lower than the FM state and becomes the magnetic ground state of this trilayer (Fig.~\ref{fig:f4}(a)). The energy maximum at the $\overline{\mathrm{K}}$ point (N\'eel state) along $\overline{\Gamma \mathrm{KM}}$ is nearly 60 meV/Fe atom higher than the FM state and the points at the BZ boundaries, i.e., both $\overline{\mathrm{M}}$ points, are slightly higher in energy than the FM state. The multi-$Q$ states remain energetically higher than the corresponding single-$Q$ states for fcc-Rh/Fe/Ir.

In contrast, the energy scale of the spin spiral dispersion of hcp-Rh/Fe/Ir at zero field (Fig.~\ref{fig:f4}(b)) is squeezed compared to that of fcc-Rh/Fe/Ir: the energy maximum at the $\overline{\mathrm{K}}$ point is reduced and the spin spiral minima become significantly shallower. Most importantly, the energy of the $uudd$ state along $\overline{\Gamma \mathrm{M}}$ is lower than the spin spiral minimum and becomes the magnetic ground state of the hcp-Rh/Fe/Ir trilayer. The $uudd$ state along $\overline{\Gamma \mathrm{KM}}$ and the 3$Q$ state at the $\overline{\mathrm{M}}$ point remain energetically higher than their corresponding single-$Q$ states.

Interestingly, the qualitative behavior of the energy dispersion and the magnetic ground state of the two trilayers remain same in the corresponding ultrathin films~\cite{romming18}. Specifically, for the two films, the spin spiral minima are observed close to the FM state along both directions, nearly 10--15 meV/Fe atom below the FM state. The energy dispersion of fcc-Rh/Fe/Ir(111) reaches a maximum value at the $\overline{\mathrm{K}}$ point along $\overline{\Gamma \mathrm{KM}}$ and at the $\overline{\mathrm{M}}$ point along $\overline{\Gamma \mathrm{M}}$, which are nearly 50--65 meV/Fe atom above the FM state. Those maxima for the energy dispersion of hcp-Rh/Fe/Ir(111) are nearly 20--40 meV/Fe atom higher than the FM state. Similar to hcp-Rh/Fe/Ir, the $uudd$ state of hcp-Rh/Fe/Ir(111) along $\overline{\Gamma \mathrm{M}}$ is energetically lower than the spin spiral minimum and is the magnetic ground state of the film. Note that the change in the magnetic ground state with stacking order of the Rh overlayer on Fe/Ir(111) was predicted by DFT and subsequently observed experimentally via spin-polarized scanning tunneling microscopy~\cite{romming18}.

The total energies of spin spiral and multi-$Q$ states of the two Rh/Fe/Ir trilayers vary in a similar way as those of the Pd/Fe/Ir trilayers under electric field, i.e., their energies reduce at $\mathcal{E}=-1.0$~V/\AA and increase at $\mathcal{E}=+1.0$~V/\AA~with respect to the zero field values (Fig.~\ref{fig:f4}). Consequently, the depth of the spin spiral minimum for
fcc-Rh/Fe/Ir changes by nearly 14 meV/Fe atom for $\mathcal{E}=\pm 1.0$~V/\AA, while the corresponding
$\textbf{q}$ value remains fixed at $0.30 \times \frac{2\pi}{a}$ (Fig.~\ref{fig:f4}(a)). For hcp-Rh/Fe/Ir, the electric-field induced modification of the energy dispersion with respect to the zero field (Fig.~\ref{fig:f4}(b)) is significantly larger than that of fcc-Rh/Fe/Ir and those of the Pd/Fe/Ir trilayers. However, the $uudd$ state of hcp-Rh/Fe/Ir along $\overline{\Gamma \mathrm{M}}$ still remains the magnetic ground state at all electric field values (Fig.~\ref{fig:f4}(b)).

To examine the trends of exchange interactions with electric fields in Rh/Fe/Ir trilayers, in addition to the above two electric fields of $\mathcal{E}=\pm 1.0$~V/\AA, we obtained the pairwise exchange and HOI constants also at intermediate fields of $\mathcal{E}=\pm 0.5$~V/\AA~(not shown) based on DFT total energy calculations of the spin spiral and multi-$Q$ states. The first three nearest-neighbor pairwise exchange constants and the three HOI constants for fcc-Rh/Fe/Ir and hcp-Rh/Fe/Ir at zero and four finite electric field values are listed in Table~\ref{tab:table2}. The pairwise exchange constants beyond third nearest neighbors are given in Table~\ref{apptab:table1} of Appendix~\ref{appsec:beyond_j3} and the energy differences between the multi-$Q$ and single-$Q$ states used to determine the HOI constants are listed in Table~\ref{apptab:table2} of Appendix~\ref{appsec:multiq-singleq}.

\begin{figure*}[!htbp]
	\includegraphics[scale=1.0]{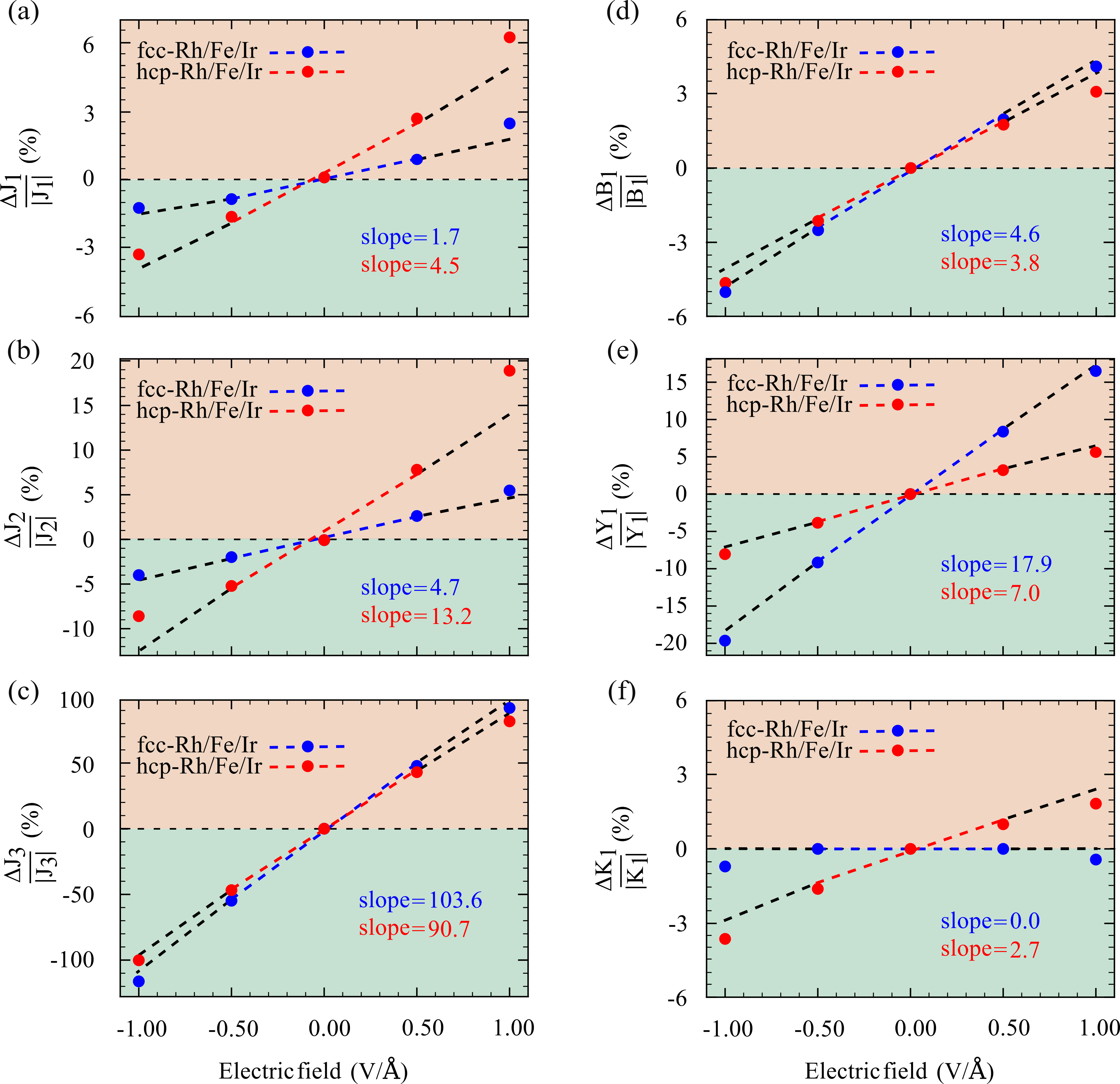}
	\centering
	\caption{\justifying Relative changes of exchange constants with electric fields for Rh/Fe/Ir trilayers. (a) first ($J_1$), (b) second ($J_2$), and (c) third ($J_3$) nearest-neighbor exchange interaction constants, and the three HOI constants, (d) biquadratic ($B_1$), (e) three-site four spin ($Y_1$), and (f) four-site four spin ($K_1$), at two positive ($0.5$~V/\AA~and $1.0$~V/\AA) and two negative ($-0.5$~V/\AA~and $-1.0$~V/\AA) electric fields with reference to the zero electric field for fcc-Rh/Fe/Ir (blue) and hcp-Rh/Fe/Ir (red). The solid circles represent DFT data. The dashed red and blue lines are obtained from least square fit up to $\mathcal{E}=\pm 0.5$~V/\AA, whereas the black dashed lines are obtained by linear extrapolations of these fits. The slope of each line is shown and given in \%/(V/\AA).} 
	\label{fig:f5}
\end{figure*}

At zero field, the sign of the first nearest-neighbor exchange constant ($J_1$) of the two Rh/Fe/Ir trilayers is positive (Table~\ref{tab:table2}). Its value is smaller by a factor of nearly 2 for fcc-Rh/Fe/Ir and 4.5 for hcp-Rh/Fe/Ir than that of the corresponding Pd/Fe/Ir trilayer, consistent with the reduced energy difference between the FM and RW-AFM states in Figs.~\ref{fig:f4}(a) and~\ref{fig:f4}(b). At zero field, the value of $J_2$ for hcp and fcc stacked Rh/Fe/Ir trilayers is on the order of the corresponding $J_1$, and it is comparable in magnitude to that of the Pd/Fe/Ir trilayer with the same stacking. Its negative sign indicates the presence of exchange frustration, which leads to stabilization of spin spiral states below the FM state in both Rh/Fe/Ir trilayers (Figs.~\ref{fig:f4}(a) and~\ref{fig:f4}(b)). The third nearest-neighbor exchange constant ($J_3$) of the Rh/Fe/ir trilayers possesses a positive sign. For hcp-Rh/Fe/Ir, its value is comparable to the other two exchange constants, while for fcc-Rh/Fe/Ir, it is one order of magnitude smaller than $J_1$ and $J_2$.

In contrast to the trilayers, the sign of $J_3$ is negative for the fcc-Rh/Fe/Ir(111) and hcp-Rh/Fe/Ir(111) ultrathin films~\cite{romming18}, while the sign of $J_1$ for both films and the sign of $J_2$ for fcc-Rh/Fe/Ir(111) remain the same as those of the trilayers. Compared to trilayers, the value of $J_1$ increases by a factor of about 2 in the films, that of $J_3$ increases by a factor of about 3--4, while that of $J_2$ decreases by a factor of nearly 0.7 in fcc-Rh/Fe/Ir(111). Surprisingly, the value of $J_2$ for the hcp-Rh/Fe/Ir(111) ultrathin film nearly vanishes.

\begin{figure*}[!htbp]
	\includegraphics[scale=1.0]{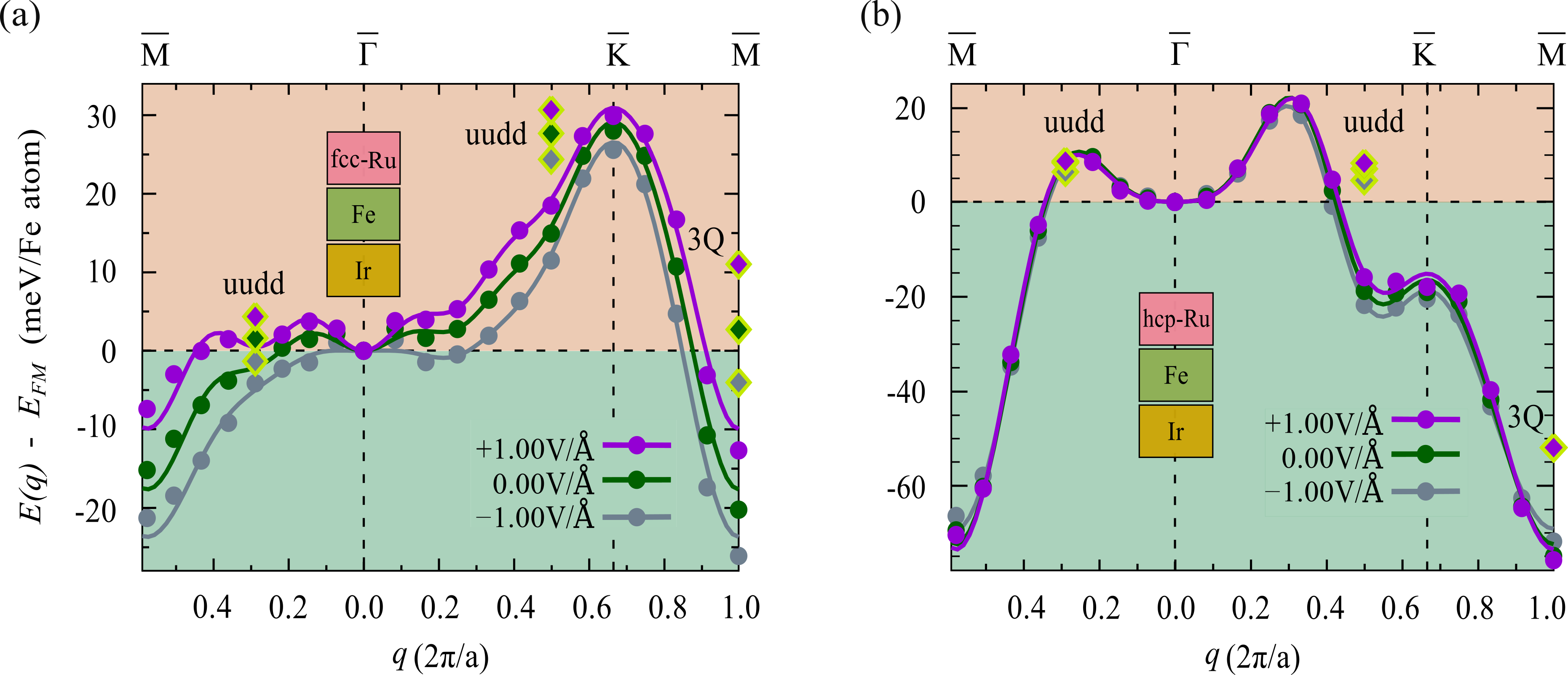}
	\centering
	\caption{\justifying Variation of spin spiral energy dispersion and multi-$Q$ states for Ru/Fe/Ir trilayers with electric fields. Energy dispersion $E$(\textbf{q}) of homogeneous flat spin spiral for (a) the fcc-Ru/Fe/Ir and (b) the hcp-Ru/Fe/Ir trilayer along the two high-symmetry directions $\overline{\Gamma \mathrm{KM}}$ and $\overline{\Gamma \mathrm{M}}$ of 2DBZ at electric fields of $\mathcal{E}$= 1.0~V/\AA~(violet), 0.0~V/\AA~(green), and $-$1.0~V/\AA~(grey).The filled circles present DFT data and solid lines are fit to the Heisenberg model (Eq.~(1)). Filled diamonds outlined by solid yellow borders represent the DFT energies of multi-$Q$ ($uudd$ and 3$Q$) states with respect to the FM state at the $q$ points of the corresponding single-$Q$ states. The energies of the multi-$Q$ states are also shown for $\mathcal{E}$= 0.0,$\pm 1.0$~V/\AA~(same color scheme as of the single-$Q$ states).} 
	\label{fig:f6}
\end{figure*}

On the other hand, the signs of all three HOI constants of the two Rh/Fe/Ir trilayers are positive at zero field (Table~\ref{tab:table2}). Their values are comparable to the exchange constants for fcc-Rh/Fe/Ir and even larger than those for hcp-Rh/Fe/Ir, indicating a stronger higher-order contribution. This stabilizes the $uudd$ state energetically below the spin spiral minimum, making it the magnetic ground state for the hcp-Rh/Fe/Ir trilayer~\cite{gutzeit2021}. The signs of the HOI constants in the corresponding films are positive, except for $Y_1$ in fcc-Rh/Fe/Ir(111), which attains a negative sign. The value of $K_1$ in both trilayers and the values of $Y_1$ and $B_1$ in hcp-Rh/Fe/Ir decrease by a factor of nearly 0.7 in the corresponding films. However, the value of $B_1$ in fcc-Rh/Fe/Ir increases by a factor of nearly 1.3 in the corresponding film, while the value of $Y_1$ in fcc-Rh/Fe/Ir remains the same in the corresponding film.

The positive signs of three HOI constants, the positive sign of $J_1$, and the negative sign of $J_2$ at zero field remain unchanged with electric fields for both Rh/Fe/Ir trilayers (Table~\ref{tab:table2}). The sign of $J_3$ also remains unchanged, i.e., positive at all electric field values, except for fcc-Rh/Fe/Ir at $\mathcal{E}= -1.0$~V/\AA, where it becomes negative. The HOI constants vary in an analogous way to those of the Pd/Fe/Ir trilayers with electric fields, i.e., their values increase (decrease) with increasing strength of the positive (negative) electric field. An exception is the four-site four spin constant ($K_1$) for fcc-Rh/Fe/Ir, whose zero field value remains unchanged with electric fields. Unlike the Pd/Fe/Ir trilayers, the values of $J_1$ and $J_3$ for the Rh/Fe/Ir trilayers vary with electric fields in a similar way to those of the HOI constants, whereas the value of $J_2$ exhibits an opposite behavior. The magnetic moment of the Fe layer at zero field for the Rh/Fe/Ir trilayers is slightly lower than that of the Pd/Fe/Ir trilayers, and the values remain almost constant under electric fields.

To examine the trends in the exchange and HOI constants of the fcc-Rh/Fe/Ir and hcp-Rh/Fe/Ir trilayers under electric fields in more detail, we display the relative changes of these interaction constants as a function of electric field in Fig.~\ref{fig:f5}. The relative changes in $J_1$ for both Rh/Fe/Ir trilayers, as well as those in $J_2$, $B_1$, and $K_1$ for hcp-Rh/Fe/Ir show a linear dependence on the electric field up to a moderate value of $0.5$~V/\AA, and deviates from linearity by small to noticeable amounts at larger electric field values. In contrast, the relative changes in $J_3$ and $Y_1$ for both Rh/Fe/Ir trilayers, as well as those in $J_2$ and $B_1$ for fcc-Rh/Fe/Ir display a linear variation with electric field up to a value of $1.0$~V/\AA.

\begin{table*} [!htbp]
	\centering
	\caption{\justifying  Variation of exchange constants of Ru/Fe/Ir trilayers with electric fields. $i$-th nearest-neighbor pairwise exchange constant ($J_{i}$), three HOI constants, i.e., biquadratic ($B_{1}$), three-site four spin ($Y_{1}$), and four-site four spin ($K_{1}$), and magnetic moment of the Fe layer ($\mu_{\mathrm{Fe}}$) in the FM state for unsupported fcc-Ru/Fe/Ir and hcp-Ru/Fe/Ir trilayers at $\mathcal{E}= 0, \pm 0.5, \pm 1.0$~V/\AA. The three nearest-neighbor exchange interaction constants are obtained from mapping the total DFT energy of spin spirals (Fig.~\ref{fig:f6}) onto the Heisenberg model (Eq.~(1)) and the HOI constants are evaluated from multi-$Q$ states using Eqs.~(5)--(10). Electric field values are given in V/\AA, magnetic interaction constants in meV, and magnetic moments in $\mu_{\mathrm{B}}$.}
	\begin{ruledtabular}
		\label{tab:table3}
		\begin{tabular}{ccccccccccccccccc}
			system & $\mathcal{E}$ & $J_{1}$ & $J_{2}$ & $J_{3}$ & $B_{1}$ & $Y_{1}$ & $K_{1}$ & $\mu_{\mathrm{Fe}}$\\
			\colrule
			& $-$1.0 & \begin{tabular}[c]{@{}c@{}}0.67\end{tabular} & \begin{tabular}[c]{@{}c@{}}$-$3.01\end{tabular} & \begin{tabular}[c]{@{}c@{}}2.72\end{tabular} & \begin{tabular}[c]{@{}c@{}}1.72\end{tabular} & \begin{tabular}[c]{@{}c@{}}1.26\end{tabular} & \begin{tabular}[c]{@{}c@{}}1.83\end{tabular} & \begin{tabular}[c]{@{}c@{}}1.76\end{tabular}\\
			
			& $-$0.5 & \begin{tabular}[c]{@{}c@{}}0.86\end{tabular} & \begin{tabular}[c]{@{}c@{}}$-$2.82\end{tabular} & \begin{tabular}[c]{@{}c@{}}2.60\end{tabular} & \begin{tabular}[c]{@{}c@{}}1.74\end{tabular} & \begin{tabular}[c]{@{}c@{}}1.25\end{tabular} & \begin{tabular}[c]{@{}c@{}}1.85\end{tabular} & \begin{tabular}[c]{@{}c@{}}1.76\end{tabular}\\
			
			fcc-Ru/Fe/Ir & 0.0 & \begin{tabular}[c]{@{}c@{}}1.10\end{tabular} & \begin{tabular}[c]{@{}c@{}}$-$2.61\end{tabular} & \begin{tabular}[c]{@{}c@{}}2.45\end{tabular} & \begin{tabular}[c]{@{}c@{}}1.79\end{tabular} & \begin{tabular}[c]{@{}c@{}}1.22\end{tabular} & \begin{tabular}[c]{@{}c@{}}1.86\end{tabular} & \begin{tabular}[c]{@{}c@{}}1.75\end{tabular}\\
			
			& $+$0.5 & \begin{tabular}[c]{@{}c@{}}1.36\end{tabular} & \begin{tabular}[c]{@{}c@{}}$-$2.40\end{tabular} & \begin{tabular}[c]{@{}c@{}}2.25\end{tabular} & \begin{tabular}[c]{@{}c@{}}1.82\end{tabular} & \begin{tabular}[c]{@{}c@{}}1.19\end{tabular} & \begin{tabular}[c]{@{}c@{}}1.86\end{tabular} & \begin{tabular}[c]{@{}c@{}}1.75\end{tabular}\\
			
			& $+$1.0 & \begin{tabular}[c]{@{}c@{}}1.64\end{tabular} & \begin{tabular}[c]{@{}c@{}}$-$2.17\end{tabular} & \begin{tabular}[c]{@{}c@{}}2.03\end{tabular} & \begin{tabular}[c]{@{}c@{}}1.84\end{tabular} & \begin{tabular}[c]{@{}c@{}}1.14\end{tabular} & \begin{tabular}[c]{@{}c@{}}1.86\end{tabular} & \begin{tabular}[c]{@{}c@{}}1.75\end{tabular}\\
			\colrule
			& \begin{tabular}[c]{@{}c@{}}$-$1.0\end{tabular} & \begin{tabular}[c]{@{}c@{}}$-$7.46\end{tabular} & \begin{tabular}[c]{@{}c@{}}0.45\end{tabular} & \begin{tabular}[c]{@{}c@{}}6.48\end{tabular} & \begin{tabular}[c]{@{}c@{}}2.11\end{tabular} & \begin{tabular}[c]{@{}c@{}}3.54\end{tabular} & \begin{tabular}[c]{@{}c@{}}2.56\end{tabular} & \begin{tabular}[c]{@{}c@{}}1.75\end{tabular}\\
			
			& \begin{tabular}[c]{@{}c@{}}$-$0.5\end{tabular} & \begin{tabular}[c]{@{}c@{}}$-$7.42\end{tabular} & \begin{tabular}[c]{@{}c@{}}0.40\end{tabular} & \begin{tabular}[c]{@{}c@{}}6.76\end{tabular} & \begin{tabular}[c]{@{}c@{}}2.14\end{tabular} & \begin{tabular}[c]{@{}c@{}}3.42\end{tabular} & \begin{tabular}[c]{@{}c@{}}2.62\end{tabular} & \begin{tabular}[c]{@{}c@{}}1.74\end{tabular}\\
			
			hcp-Ru/Fe/Ir & \begin{tabular}[c]{@{}c@{}}0.0\end{tabular} & \begin{tabular}[c]{@{}c@{}}$-$7.33\end{tabular} & \begin{tabular}[c]{@{}c@{}}0.36\end{tabular} & \begin{tabular}[c]{@{}c@{}}6.95\end{tabular} & \begin{tabular}[c]{@{}c@{}}2.15\end{tabular} & \begin{tabular}[c]{@{}c@{}}3.28\end{tabular} & \begin{tabular}[c]{@{}c@{}}2.64\end{tabular} & \begin{tabular}[c]{@{}c@{}}1.74\end{tabular}\\
			
			& \begin{tabular}[c]{@{}c@{}}$+$0.5\end{tabular} & \begin{tabular}[c]{@{}c@{}}$-$7.18\end{tabular} & \begin{tabular}[c]{@{}c@{}}0.31\end{tabular} & \begin{tabular}[c]{@{}c@{}}7.06\end{tabular} & \begin{tabular}[c]{@{}c@{}}2.17\end{tabular} & \begin{tabular}[c]{@{}c@{}}3.12\end{tabular} & \begin{tabular}[c]{@{}c@{}}2.63\end{tabular} & \begin{tabular}[c]{@{}c@{}}1.73\end{tabular}\\
			
			& \begin{tabular}[c]{@{}c@{}}$+$1.0\end{tabular} & \begin{tabular}[c]{@{}c@{}}$-$6.98\end{tabular} & \begin{tabular}[c]{@{}c@{}}0.27\end{tabular} & \begin{tabular}[c]{@{}c@{}}7.09\end{tabular} & \begin{tabular}[c]{@{}c@{}}2.17\end{tabular} & \begin{tabular}[c]{@{}c@{}}2.96\end{tabular} & \begin{tabular}[c]{@{}c@{}}2.60\end{tabular} & \begin{tabular}[c]{@{}c@{}}1.73\end{tabular}\\
			
			\end{tabular}
	\end{ruledtabular}
\end{table*}
\begin{figure*}[!htbp]
	\includegraphics[scale=1.0]{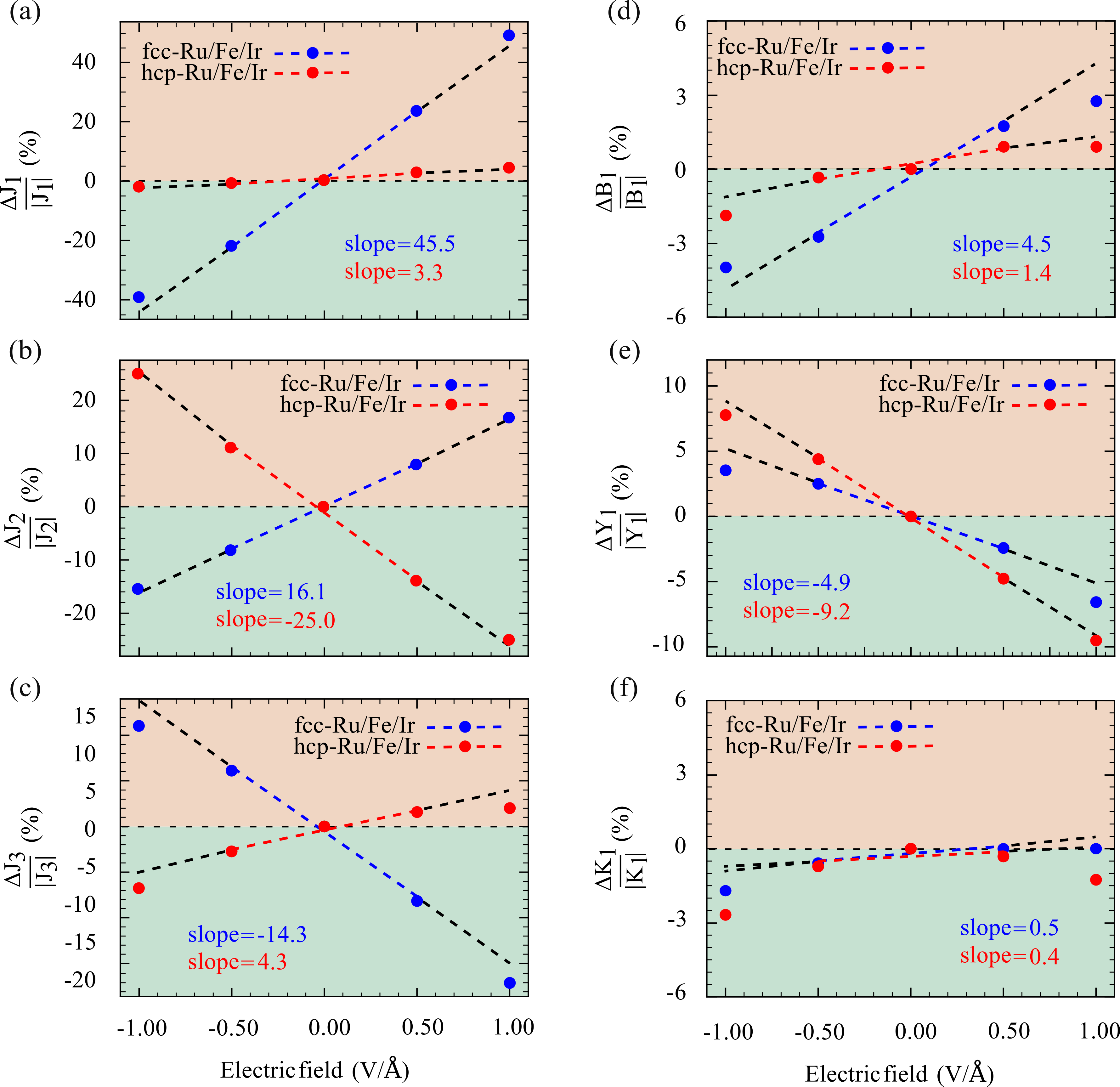}
	\centering
	\caption{\justifying Relative changes of exchange constants with electric fields for Ru/Fe/Ir trilayers. (a) first ($J_1$), (b) second ($J_2$), and (c) third ($J_3$) nearest-neighbor exchange interaction constants, and three HOI constants, (d) biquadratic ($B_1$), (e) three-site four spin ($Y_1$), and (f) four-site four spin ($K_1$) interactions, at two positive ($0.5$~V/\AA~and $1.0$~V/\AA) and two negative ($-0.5$~V/\AA~and $-1.0$~V/\AA) electric fields with reference to the zero electric field for fcc-Ru/Fe/Ir (blue) and hcp-Ru/Fe/Ir (red). The solid circles represent DFT data. The dashed red and blue lines are obtained from least square fit up to $\mathcal{E}=\pm 0.5$~V/\AA, whereas the black dashed lines are obtained by linear extrapolations of these fits. The slope of each line is shown and given in \%/(V/\AA).} 
	\label{fig:f7}
\end{figure*} 

The slopes of all constants for both Rh/Fe/Ir trilayers are positive (blue symbols and lines in Fig.~\ref{fig:f5}(f)), except for $K_1$ in fcc-Rh/Fe/Ir, whose slope is zero since its value remains constant at all electric field values (Table~\ref{tab:table2}). Interestingly, the changes in $J_3$ for both Rh/Fe/Ir trilayers are one order of magnitude greater than the largest changes in the Pd/Fe/Ir trilayers, resulting in a slope magnitude on the order of 100~\%/(V/\AA). The slopes of $J_2$ for hcp-Rh/Fe/Ir and $Y_1$ for fcc-Rh/Fe/Ir exhibit moderate values, between 13 and 17~\%/(V/\AA), while the slope magnitude of $Y_1$ for hcp-Rh/Fe/Ir is 7~\%/(V/\AA). The values of the slopes for all other constants remain below 5~\%/(V/\AA).

Overall, we find that the electric field modifies the energy dispersion of spin spiral and the energies of multi-$Q$ states of the Rh/Fe/Ir trilayers in a similar way as those of the Pd/Fe/Ir trilayers. The field induced changes in several magnetic interaction constants follow a linear relation over the entire range of electric fields, whereas some constants deviate by small to substantial amounts at large electric field values.

\subsection{Ru/Fe/Ir trilayers}

Next, we choose a Ru overlayer in place of Rh to examine the effect of band filling and study the Ru/Fe/Ir trilayer. Both fcc and hcp stacking of the Ru overlayer are considered for comparison with the Pd/Fe/Ir and Rh/Fe/Ir trilayers.

The spin spiral energy dispersion along the $\overline{\Gamma \mathrm{KM}}$ and $\overline{\Gamma\mathrm{M}}$ directions of 2DBZ at zero field for the fcc-Ru/Fe/Ir and hcp-Ru/Fe/Ir trilayers (Figs.~\ref{fig:f6}(a) and \ref{fig:f6}(b)) is completely different from that of the Pd/Fe/Ir and Rh/Fe/Ir trilayers and the RW-AFM configuration becomes the magnetic ground state.

\begin{figure*}[!htbp]
	\includegraphics[scale=1.0]{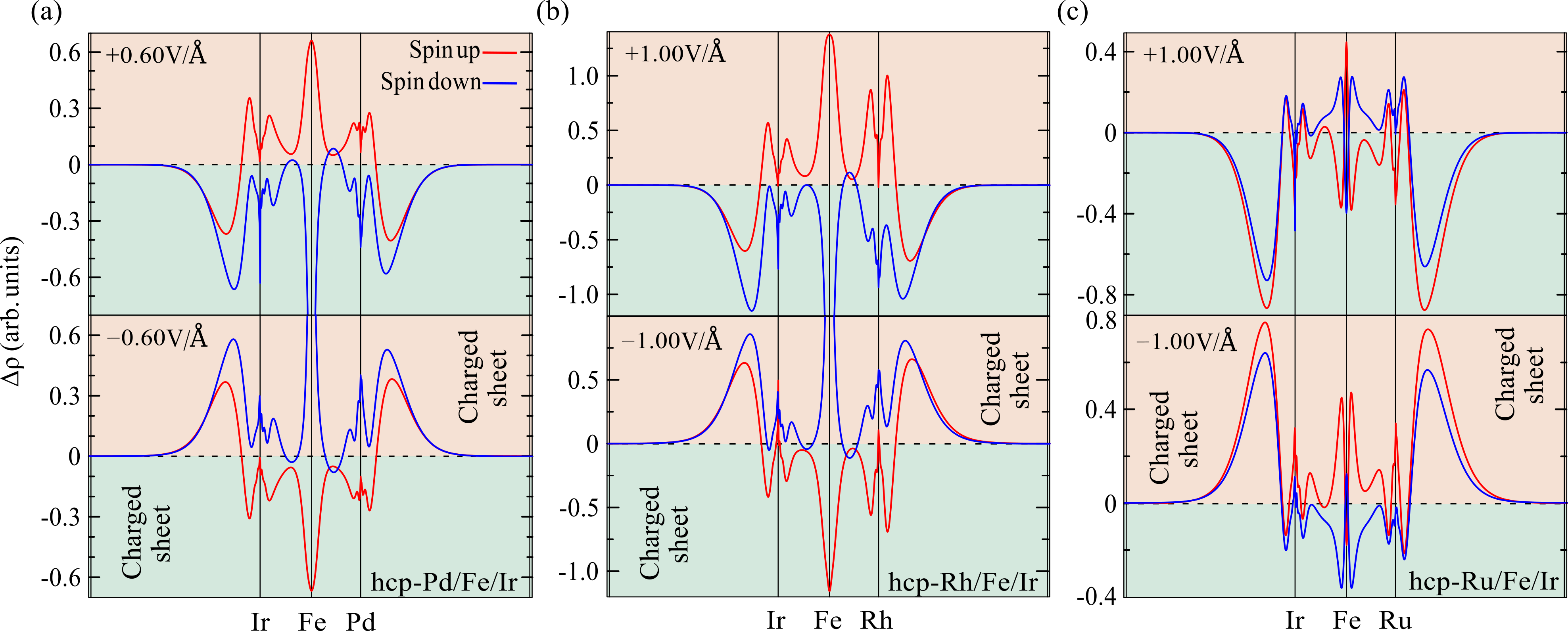}
	\centering
	\caption{\justifying Spin-dependent screening of the electric field. Spin-dependent charge density difference ($\Delta \rho_{\sigma} (z,\mathcal{E})= \rho_{\sigma} (z,\mathcal{E}) - \rho_{\sigma} (z,0)$) for (a) hcp-Pd/Fe/Ir at $\mathcal{E} = \pm 0.6$~V/\AA~and (b) hcp-Rh/Fe/Ir and (c) hcp-Ru/Fe/Ir at $\mathcal{E} = \pm 1.0$~V/\AA~with reference to $\mathcal{E} = 0.0$~V/\AA~along the $z$-direction of the trilayer, while being averaged over the film plane ($xy$). The top (bottom) panels show $\Delta \rho (z,\mathcal{E})$ for positive (negative) electric field. The spin-up ($\sigma$ = $\uparrow$) and spin-down ($\sigma$ = $\downarrow$) charge density difference is denoted by the solid red and blue line, respectively. The position of the charged sheets, as well as the positions of the 4$d$ (Pd, Rh, and Ru), Fe, and Ir layers are also indicated.}
	\label{fig:f8}
\end{figure*}

The zero field energy dispersion of fcc-Ru/Fe/Ir is consistent with that reported in Ref.~\cite{gutzeit2021}. It rises from the FM state ($\overline{\Gamma}$ point) with an increasing value of the wave vector $\textbf{q}$ along $\overline{\Gamma \mathrm{KM}}$, reaches a maximum of nearly 30 meV/Fe atom higher than the FM state at the $\overline{\mathrm{K}}$ point, signifying the N\'eel state (Fig.~\ref{fig:f6}(a)). The dispersion drops sharply after this point and attains a minimum value of nearly 20 meV/Fe atom below the FM state at the BZ boundary ($\overline{\mathrm{M}}$ point), representing the RW-AFM state. The dispersion along $\overline{\Gamma \mathrm{M}}$ drops slowly from the FM state with increasing $q$ value and attains nearly the same minimum energy at the BZ boundary ($\overline{\mathrm{M}}$ point). Therefore, the RW-AFM configuration becomes the magnetic ground state of this trilayer. The energies of the multi-$Q$ states ($uudd$ and $3Q$) remain higher than their corresponding single-$Q$ states, but the energy difference between the multi-$Q$ and single-$Q$ states along $\overline{\Gamma \mathrm{M}}$ is noticeably smaller than those in the other direction.

The zero field energy dispersion of hcp-Ru/Fe/Ir (Fig.~\ref{fig:f6}(b)) first rises slowly from the FM state with increasing $q$ value along both directions and reaches a maximum value of nearly 20 meV/Fe atom higher than the FM state at $q=0.30 \times \frac{2\pi}{a}$ along $\overline{\Gamma \mathrm{KM}}$ and nearly 10 meV/Fe atom higher than the FM state at $q=0.28 \times \frac{2\pi}{a}$ along $\overline{\Gamma \mathrm{M}}$. Thereafter, the dispersion drops with further increase in $q$ along both directions and reaches a minimum at the zone boundary of nearly 70 meV/Fe atom below the FM state, representing the RW-AFM state, which is the magnetic ground state of this trilayer. The total energies of the multi-$Q$ states ($uudd$ and $3Q$) along $\overline{\Gamma \mathrm{KM}}$ are noticeably higher than those of the corresponding single-$Q$ states, while the $uudd$ state along $\overline{\Gamma \mathrm{M}}$ is almost degenerate in energy with its corresponding single-$Q$ states.

Note that the energy dispersions of Ru/Fe/Ir trilayers shown in Figs.~\ref{fig:f6}(a) and~\ref{fig:f6}(b) are modified by $\frac{1}{2}I\left(M(\textbf{q})-M(0)\right)^2$, where $M(\textbf{q})$ is the magnetic moment obtained via DFT for a spin spiral vector $\textbf{q}$ and $I$ is the Stoner parameter fixed at 420 meV for Fe, as suggested in Ref.~\cite{gutzeit2021}. This correction accounts for the large variation in magnetic moments ($M(\textbf{q})$) across the two high-symmetry directions of 2DBZ (not shown) and allows us to map the spin spiral energies onto the Heisenberg Hamiltonian (Eq.~(1)) for the extraction of the pairwise exchange constants.

The energy dispersion of spin spirals and energies of multi-$Q$ states for fcc-Ru/Fe/Ir and hcp-Ru/Fe/Ir at zero field vary in a similar manner with electric fields as those of the Pd/Fe/Ir and Rh/Fe/Ir trilayers (Figs.~\ref{fig:f6}(a) and \ref{fig:f6}(b)). However, the electric field induced changes are significantly smaller for the hcp-Ru/Fe/Ir trilayer compared to fcc-Ru/Fe/Ir. Nevertheless, the RW-AFM state remains the magnetic ground state of these two trilayers at finite electric field values.

The first three nearest-neighbor pairwise exchange constants and the three HOI constants are reported for the fcc-Ru/Fe/Ir and hcp-Ru/Fe/Ir trilayers at $\mathcal{E}= 0, \pm 0.5, \pm 1.0$~V/\AA~in Table~\ref{tab:table3}. The beyond third nearest-neighbor exchange constants and the energy differences between the multi-$Q$ and single-$Q$ states used to determine the HOI constants for these two Ru/Fe/Ir trilayers at the above mentioned field values are listed in Table~\ref{apptab:table1} of Appendix~\ref{appsec:beyond_j3} and Table~\ref{apptab:table2} of Appendix~\ref{appsec:multiq-singleq}, respectively.

The sign of the first nearest-neighbor exchange constant ($J_1$) for fcc-Ru/Fe/Ir at zero field is positive, favoring the FM coupling. However, its value, $\approx$ 1 meV, is significantly lower than that of the Pd/Fe/Ir and Rh/Fe/Ir trilayers (Table~\ref{tab:table3}). The sign of the third nearest-neighbor exchange constant ($J_3$) is also positive, and its magnitude is larger than $J_1$ by a factor of nearly 2. In contrast, the sign of the second nearest-neighbor constant ($J_2$) is negative, favoring the AFM coupling, and its magnitude is larger than $J_1$ by a factor of nearly 2. These competing interactions induce exchange frustration in the trilayer and stabilizes the RW-AFM state below the FM states, as shown in Fig.~\ref{fig:f6}(a).

On the other hand, the sign of $J_1$ for hcp-Ru/Fe/Ir is negative at zero field and its value is approximately 7 meV, which strongly favors the AFM coupling (Table~\ref{tab:table3}). Both $J_2$ and $J_3$ possess positive signs and therefore, favor FM coupling. The value of $J_3$ is comparable to $J_1$, while that of $J_2$ is one order of magnitude smaller. These competing FM and AFM interactions stabilize the RW-AFM state below the FM state, which becomes the ground state of this system. The signs of the HOI constants at zero field are positive for both Ru/Fe/Ir trilayers and their values are comparable to the exchange constants.

The signs of all interaction constants at zero field remain unchanged at the chosen electric field values for the two Ru/Fe/Ir trilayers (Table~\ref{tab:table3}). The values of $J_2$, $B_1$, and $Y_1$ vary in a similar way with the electric field for the two Ru/Fe/Ir trilayers, while the values of $J_1$ and $J_3$ display an opposite trend between the two trilayers. Specifically, the values of $J_2$ and $Y_1$ increase (decrease) with increasing strength of the negative (positive) electric field, while the value of $B_1$ exhibits an opposite trend with the electric field. The variation in the value of $J_1$ for fcc-Ru/Fe/Ir and $J_3$ for hcp-Ru/Fe/Ir follows the same trend as $B_1$, while that of $J_1$ for hcp-Ru/Fe/Ir and $J_3$ for fcc-Ru/Fe/Ir follows the same trend as $J_2$. The zero field value of $K_1$ remains almost constant with electric field for both Ru/Fe/Ir trilayers. The magnetic moment of the Fe layer at zero field for the Ru/Fe/Ir trilayers is lower than the Pd/Fe/Ir and Rh/Fe/Ir trilayers by factors of nearly 1.6 and 1.4, respectively, and the values remain almost constant under the electric field.

The relative changes of all six interaction constants for the fcc-Ru/Fe/Ir and hcp-Ru/Fe/Ir trilayers with electric field are shown in Fig.~\ref{fig:f7}. Similar to the Pd/Fe/Ir and Rh/Fe/Ir trilayers, all constants of the two Ru/Fe/Ir trilayers exhibit a linear variation with the electric field up to a moderate value of $\mathcal{E}= \pm 0.5$~V/\AA, and the field induced changes in all constants, except $J_2$, deviate from the linear behavior at higher electric field values. Only the variation of $J_2$ in both trilayers follows a linear relation with the electric field up to $\mathcal{E}= \pm 1.0$~V/\AA.

\begin{table*}[!htbp]
	\centering
	\caption{\justifying Variation of orbital charge for hcp-Rh/Fe/Ir with electric fields. Difference in electron occupation of Rh 5$s$, 5$p$, and 4$d$, Fe 4$s$, 4$p$, and 3$d$, and Ir 6$s$, 6$p$, and 5$d$ valence orbitals at $\mathcal{E} = \pm 1.0$~V/\AA~with reference to $\mathcal{E} = 0.0$~V/\AA. The electronic occupation is shown for spin-up (majority) and spin-down (minority) channels and it is calculated from the integrated charge within the muffin-tin sphere. The electric field strength $\mathcal{E}$ is expressed in V/\AA.}
    \label{tab:table4}
\begin{ruledtabular}
\begin{tabular}{cc ccc ccc ccc}

& &
\multicolumn{3}{c}{Rh ($\times10^{-3}$)} &
\multicolumn{3}{c}{Fe ($\times10^{-3}$)} &
\multicolumn{3}{c}{Ir ($\times10^{-3}$)} \\
\cline{3-5}
\cline{6-8}
\cline{9-11}
 $\mathcal{E}$ & spin &
 5$s$ & 5$p$ & 4$d$ &
 4$s$ & 4$p$ & 3$d$ &
 6$s$ & 6$p$ & 5$d$ \\

\midrule

1.0 & up   & $-$0.4 & $-$1.0 & 13.8 & 0.4 & 0.8 & 17.3 & $-$0.02 & $-$0.6 & 6.7 \\
1.0 & down & $-$0.9 & $-$1.3 & $-$12.7 & $-$0.4 & 1.1 & $-$22.6 & $-$0.5 & $-$2.2 & $-$5.4 \\
\colrule
$-$1.0 & up  & 0.2 & 0.9 & $-$8.2  & $-$0.3 & $-$0.7 & $-$12.4 & $-$0.03 & 0.7 & $-$3.3 \\
$-$1.0 & down& 0.4 & 1.1 & 7.6 & 0.2 & $-$0.9 & 16.9 & 0.1 & 1.4 & 3.7 \\

\end{tabular}
\end{ruledtabular}
\end{table*}

The slope corresponding to $B_1$ is positive for both Ru/Fe/Ir trilayers, while it is negative for $Y_1$ (Fig.~\ref{fig:f7}). These slopes vary from small to moderate values, between 1 and 9~\%/(V/\AA). The slope associated with $K_1$ is positive for both trilayers, but the values are significantly small, nearly 0.5\%/(V/\AA), consistent with the values in Table~\ref{tab:table3}. The slope of $J_2$ is positive for fcc-Ru/Fe/Ir and negative for hcp-Ru/Fe/Ir, but the values are comparable, i.e., 16 and 25~\%/(V/\AA), respectively. The slope of $J_3$ changes sign between fcc-Ru/Fe/Ir and hcp-Ru/Fe/Ir, while that of $J_1$ remains positive for both Ru/Fe/Ir trilayers. However, the values of the slopes correspond to these constants change by one order of magnitude between the two trilayers.

From the above analysis, we observe that the electric field modifies the collinear and noncollinear states of the three series of trilayers, Pd/Fe/Ir, Rh/Fe/Ir, Ru/Fe/Ir, with fcc and hcp stackings of the overlayer, in a qualitatively similar way. The magnitude of these changes vary from small to large amounts, which is also reflected in the calculated pairwise exchange and HOI constants. Interestingly, variation of these magnetic interaction constants for all trilayers displays a linear relationship up to the moderate electric field values, and it deviates from linearity at higher electric field values for the Rh/Fe/Ir and Ru/Fe/Ir trilayers.

\subsection{ Spin-dependent screening of the electric field}
\label{subsec:spin_screening}

In order to gain insight into the spin-dependent screening of the electric field in the trilayers, we study the field induced modification of the spin-resolved charge density and local density of states (LDOS). The spin-resolved charge density at maximum positive and negative electric field values, i.e., at $\mathcal{E} = \pm 0.6$~V/\AA~for the hcp-Pd/Fe/Ir trilayer and at $\mathcal{E} = \pm 1.0$~V/\AA~for the hcp-Rh/Fe/Ir and hcp-Ru/Fe/Ir trilayers, with reference to the zero field value, are shown in Figs.~\ref{fig:f8}(a)--\ref{fig:f8}(c), respectively.

\begin{figure*}[!htbp]
	\includegraphics[scale=1.0]{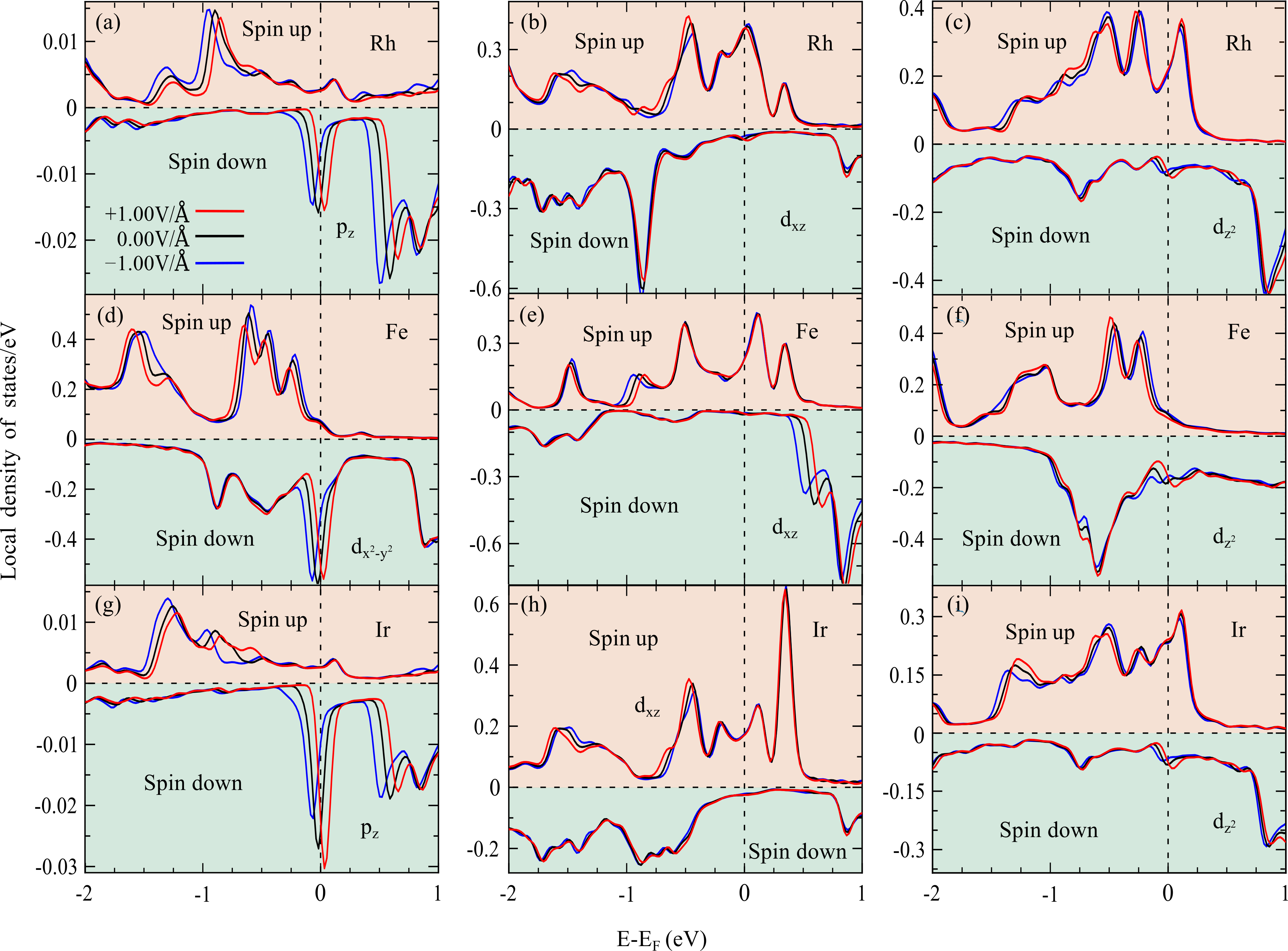}
	\centering
	\caption{\justifying Spin-resolved local density of states (LDOS) of hcp-Rh/Fe/Ir under electric fields. The top panels, (a)--(c), show LDOS of Rh $p_{z}$, $d_{xz}$, and $d_{z^2}$ orbitals, respectively. The middle panels, (d)--(f), show LDOS of Fe $d_{x^2-y^2}$, $d_{xz}$, and $d_{z^2}$ orbitals, respectively. The bottom panels, (g)--(i), show LDOS of Ir $p_{z}$, $d_{xz}$, and $d_{z^2}$ orbitals, respectively. Each panel shows LDOS at $\mathcal{E} = 1.0$~V/\AA~(red), $0.0$~V/\AA~(black), and $-1.0$~V/\AA~(blue). The Fermi energy is set to zero.}
	\label{fig:f9}
\end{figure*}

In case of a positive electric field (top panels of Figs.~\ref{fig:f8}(a)--\ref{fig:f8}(c)), the field lines point from the top and bottom layers toward the charged sheets, driving electrons toward the Fe layer. The resulting charge separation creates interfacial electric dipoles which screen the external electric field. The surface electric dipole reverses its direction as the applied field switches its polarity (bottom panels of Figs.~\ref{fig:f8}(a)--\ref{fig:f8}(c)).

The modification in the charge density due to redistribution of electrons in response to the external electric field is strongly spin-dependent within the hcp-Pd/Fe/Ir and hcp-Rh/Fe/Ir trilayers, while it is considerably less spin-dependent in front of these two trilayers, i.e., in the vacuum region (Figs.~\ref{fig:f8}(a) and~\ref{fig:f8}(b)). In contrast, the charge density difference is nearly spin-independent close to and at the top and bottom layers of hcp-Ru/Fe/Ir (Fig.~\ref{fig:f8}(c)).

The charge density difference is sizable in the vacuum, i.e., just above the top and bottom layers, as well as at all three atomic layers for hcp-Pd/Fe/Ir and hcp-Rh/Fe/Ir (Figs.~\ref{fig:f8}(a) and \ref{fig:f8}(b)). The difference is also substantial in the vacuum for hcp-Ru/Fe/Ir, but it reduces significantly at the three atomic layers (Fig.~\ref{fig:f8}(c)). 

To explain the spin-dependent variation of the charge density under the electric field, we present the spin-resolved LDOS of selected orbitals for all three atomic layers of the hcp-Rh/Fe/Ir trilayer in Fig.~\ref{fig:f9}. The large charge density difference in the vacuum, i.e., in front of the top (Rh) and bottom (Ir) layers, originates mainly from the $p_{z}$ orbital of the Rh (Fig.~\ref{fig:f9}(a)) and Ir (Fig.~\ref{fig:f9}(g)) atoms, which extend far into the vacuum.

The prominent spin-minority, i.e., spin-down, peak of the $p_{z}$ orbital at the Fermi energy for both Rh (Fig.~\ref{fig:f9}(a)) and Ir (Fig.~\ref{fig:f9}(g)) atoms in the absence of an electric field shifts to a higher (lower) energy upon application of the positive (negative) electric field. A similar shift is observed for the less prominent spin-minority peak at the Fermi level of $d_{z^2}$ character for Rh and Ir (Figs.~\ref{fig:f9}(c) and~\ref{fig:f9}(i)). These shifts indicate a decrease (increase) of spin-down electrons in the $p_{z}$ and $d_{z^2}$ orbitals at $\mathcal{E} = 1.0$~V/\AA~($\mathcal{E} = -1.0$~V/\AA). The zero-field spin-majority, i.e., spin-up, peak of the $p_{z}$ orbital for Rh at around 1 eV below the Fermi energy and for Ir at around 1.3 eV below the Fermi energy decreases (increases) in height at the positive (negative) electric field, leading to a similar change as that of the occupied $p_{z}$ states. These modifications of the $p_{z}$ and $d_{z^2}$ orbitals explain the spin-dependent charge density difference observed in the vacuum of Fig.~\ref{fig:f8}(b).

The changes in the spin-resolved LDOS of the $p_{z}$ states discussed above are supported by the modifications in the spin-dependent integrated charges of the Rh and Ir $p$ orbitals within the muffin-tin spheres (Table~\ref{tab:table4}). The charge of the Rh 5$p$ orbitals in the spin-up and spin-down channels decreases (increases) by 1.0 and 1.3$\times 10^{-3}$ electrons, respectively (0.9 and 1.1$\times 10^{-3}$ electrons, respectively) at $\mathcal{E} = 1.0$~V/\AA~($\mathcal{E} = -1.0$~V/\AA) with reference to the zero field value. Similarly, the charge of the Ir 6$p$ orbitals in both spin channels decreases (increases) by 0.6 and 2.2$\times 10^{-3}$ electrons (0.7 and 1.4$\times 10^{-3}$ electrons) at $\mathcal{E} = 1.0$~V/\AA~($\mathcal{E} = -1.0$~V/\AA). These variations of electronic occupations in the $p$ orbitals explain the behavior of the charge density in front of the Rh and Ir atomic layers (Fig.~\ref{fig:f8}(b)).

\begin{table} [!htbp]
	\centering
	\caption{\justifying Variation of exchange constants for the six trilayers with electric fields. $i$-th nearest-neighbor pairwise exchange constant ($J_{i}$) for fcc-Pd/Fe/Ir and hcp-Pd/Fe/Ir trilayers at $\mathcal{E}=0,\pm 0.3,\pm 0.6$~V/\AA, and fcc-Rh/Fe/Ir, hcp-Rh/Fe/Ir, fcc-Ru/Fe/Ir, and hcp-Ru/Fe/Ir trilayers at $\mathcal{E}= 0, \pm 0.5, \pm 1.0$~V/\AA. The beyond third nearest-neighbor exchange interaction constants are calculated from mapping the total DFT energy of spin spirals (Figs.~\ref{fig:f2}, \ref{fig:f4}, and \ref{fig:f6} onto the Heisenberg model (Eq.~(1)). Electric field values are given in V/\AA~and magnetic interaction constants in meV.}
	\begin{ruledtabular}
		\begin{tabular}{ccccccc}
			system & $\mathcal{E}$ & $J_{4}$ & $J_{5}$ & $J_{6}$ & $J_{7}$ & $J_{8}$ \\
			\colrule
			& \begin{tabular}[c]{@{}c@{}}$-$0.6\end{tabular} & \begin{tabular}[c]{@{}c@{}}0.20\end{tabular} & \begin{tabular}[c]{@{}c@{}}1.17\end{tabular} & \begin{tabular}[c]{@{}c@{}}0.10\end{tabular} & \begin{tabular}[c]{@{}c@{}}0.10\end{tabular} & \begin{tabular}[c]{@{}c@{}}$-$0.12\end{tabular}\\
			
			& \begin{tabular}[c]{@{}c@{}}$-$0.3\end{tabular} & \begin{tabular}[c]{@{}c@{}}0.18\end{tabular} & \begin{tabular}[c]{@{}c@{}}1.20\end{tabular} & \begin{tabular}[c]{@{}c@{}}0.13\end{tabular} & \begin{tabular}[c]{@{}c@{}}0.12\end{tabular} & \begin{tabular}[c]{@{}c@{}}$-$0.08\end{tabular}\\
			
			fcc-Pd/Fe/Ir & \begin{tabular}[c]{@{}c@{}}0.0\end{tabular} & \begin{tabular}[c]{@{}c@{}}0.17\end{tabular} & \begin{tabular}[c]{@{}c@{}}1.23\end{tabular} & \begin{tabular}[c]{@{}c@{}}0.15\end{tabular} & \begin{tabular}[c]{@{}c@{}}0.13\end{tabular} & \begin{tabular}[c]{@{}c@{}}$-$0.05\end{tabular}\\
			
			& \begin{tabular}[c]{@{}c@{}}$+$0.3\end{tabular} & \begin{tabular}[c]{@{}c@{}}0.15\end{tabular} & \begin{tabular}[c]{@{}c@{}}1.26\end{tabular} & \begin{tabular}[c]{@{}c@{}}0.18\end{tabular} & \begin{tabular}[c]{@{}c@{}}0.14\end{tabular} & \begin{tabular}[c]{@{}c@{}}$-$0.02\end{tabular}\\
			
			& \begin{tabular}[c]{@{}c@{}}$+$0.6\end{tabular} & \begin{tabular}[c]{@{}c@{}}0.13\end{tabular} & \begin{tabular}[c]{@{}c@{}}1.29\end{tabular} & \begin{tabular}[c]{@{}c@{}}0.20\end{tabular} & \begin{tabular}[c]{@{}c@{}}0.16\end{tabular} & \begin{tabular}[c]{@{}c@{}}$-$0.01\end{tabular}\\
			\colrule
			& \begin{tabular}[c]{@{}c@{}}$-$0.6\end{tabular} & \begin{tabular}[c]{@{}c@{}}0.33\end{tabular} & \begin{tabular}[c]{@{}c@{}}0.80\end{tabular} & \begin{tabular}[c]{@{}c@{}}$-$0.22\end{tabular} & \begin{tabular}[c]{@{}c@{}}0.11\end{tabular} & \begin{tabular}[c]{@{}c@{}}$-$0.30\end{tabular}\\
			
			& \begin{tabular}[c]{@{}c@{}}$-$0.3\end{tabular} & \begin{tabular}[c]{@{}c@{}}0.31\end{tabular} & \begin{tabular}[c]{@{}c@{}}0.82\end{tabular} & \begin{tabular}[c]{@{}c@{}}$-$0.22\end{tabular} & \begin{tabular}[c]{@{}c@{}}0.11\end{tabular} & \begin{tabular}[c]{@{}c@{}}$-$0.28\end{tabular}\\
			
			hcp-Pd/Fe/Ir & \begin{tabular}[c]{@{}c@{}}0.0\end{tabular} & \begin{tabular}[c]{@{}c@{}}0.30\end{tabular} & \begin{tabular}[c]{@{}c@{}}0.84\end{tabular} & \begin{tabular}[c]{@{}c@{}}$-$0.22\end{tabular} & \begin{tabular}[c]{@{}c@{}}0.12\end{tabular} & \begin{tabular}[c]{@{}c@{}}$-$0.26\end{tabular}\\
			
			& \begin{tabular}[c]{@{}c@{}}$+$0.3\end{tabular} & \begin{tabular}[c]{@{}c@{}}0.30\end{tabular} & \begin{tabular}[c]{@{}c@{}}0.86\end{tabular} & \begin{tabular}[c]{@{}c@{}}$-$0.21\end{tabular} & \begin{tabular}[c]{@{}c@{}}0.14\end{tabular} & \begin{tabular}[c]{@{}c@{}}$-$0.23\end{tabular}\\
			
			& \begin{tabular}[c]{@{}c@{}}$+$0.6\end{tabular} & \begin{tabular}[c]{@{}c@{}}0.31\end{tabular} & \begin{tabular}[c]{@{}c@{}}0.89\end{tabular} & \begin{tabular}[c]{@{}c@{}}$-$0.20\end{tabular} & \begin{tabular}[c]{@{}c@{}}0.15\end{tabular} & \begin{tabular}[c]{@{}c@{}}$-$0.19\end{tabular}\\
			\colrule
			& \begin{tabular}[c]{@{}c@{}}$-$1.0\end{tabular} & \begin{tabular}[c]{@{}c@{}}$-$0.41\end{tabular} & \begin{tabular}[c]{@{}c@{}}0.63\end{tabular} & \begin{tabular}[c]{@{}c@{}}$-$0.37\end{tabular} & \begin{tabular}[c]{@{}c@{}}0.20\end{tabular} & \begin{tabular}[c]{@{}c@{}}$-$0.10\end{tabular}\\
			
			& \begin{tabular}[c]{@{}c@{}}$-$0.5\end{tabular} & \begin{tabular}[c]{@{}c@{}}$-$0.42\end{tabular} & \begin{tabular}[c]{@{}c@{}}0.61\end{tabular} & \begin{tabular}[c]{@{}c@{}}$-$0.36\end{tabular} & \begin{tabular}[c]{@{}c@{}}0.21\end{tabular} & \begin{tabular}[c]{@{}c@{}}$-$0.07\end{tabular}\\
			
			fcc-Rh/Fe/Ir & \begin{tabular}[c]{@{}c@{}}0.0\end{tabular} & \begin{tabular}[c]{@{}c@{}}$-$0.44\end{tabular} & \begin{tabular}[c]{@{}c@{}}0.59\end{tabular} & \begin{tabular}[c]{@{}c@{}}$-$0.34\end{tabular} & \begin{tabular}[c]{@{}c@{}}0.21\end{tabular} & \begin{tabular}[c]{@{}c@{}}$-$0.04\end{tabular}\\
			
			& \begin{tabular}[c]{@{}c@{}}$+$0.50\end{tabular} & \begin{tabular}[c]{@{}c@{}}$-$0.45\end{tabular} & \begin{tabular}[c]{@{}c@{}}0.58\end{tabular} & \begin{tabular}[c]{@{}c@{}}$-$0.31\end{tabular} & \begin{tabular}[c]{@{}c@{}}0.21\end{tabular} & \begin{tabular}[c]{@{}c@{}}$-$0.01\end{tabular}\\
			
			& \begin{tabular}[c]{@{}c@{}}$+$1.0\end{tabular} & \begin{tabular}[c]{@{}c@{}}$-$0.47\end{tabular} & \begin{tabular}[c]{@{}c@{}}0.57\end{tabular} & \begin{tabular}[c]{@{}c@{}}$-$0.28\end{tabular} & \begin{tabular}[c]{@{}c@{}}0.20\end{tabular} & \begin{tabular}[c]{@{}c@{}}$-$0.03\end{tabular}\\
			
			\colrule
			& \begin{tabular}[c]{@{}c@{}}$-$1.0\end{tabular} & \begin{tabular}[c]{@{}c@{}}0.11\end{tabular} & \begin{tabular}[c]{@{}c@{}}0.29\end{tabular} & \begin{tabular}[c]{@{}c@{}}$-$0.74\end{tabular} & \begin{tabular}[c]{@{}c@{}}$-$0.02\end{tabular} & \begin{tabular}[c]{@{}c@{}}$-$0.39\end{tabular}\\
			
			& \begin{tabular}[c]{@{}c@{}}$-$0.5\end{tabular} & \begin{tabular}[c]{@{}c@{}}0.09\end{tabular} & \begin{tabular}[c]{@{}c@{}}0.28\end{tabular} & \begin{tabular}[c]{@{}c@{}}$-$0.83\end{tabular} & \begin{tabular}[c]{@{}c@{}}$-$0.06\end{tabular} & \begin{tabular}[c]{@{}c@{}}$-$0.42\end{tabular}\\
			
			hcp-Rh/Fe/Ir & \begin{tabular}[c]{@{}c@{}}0.0\end{tabular} & \begin{tabular}[c]{@{}c@{}}0.08\end{tabular} & \begin{tabular}[c]{@{}c@{}}0.27\end{tabular} & \begin{tabular}[c]{@{}c@{}}$-$0.91\end{tabular} & \begin{tabular}[c]{@{}c@{}}$-$0.10\end{tabular} & \begin{tabular}[c]{@{}c@{}}$-$0.43\end{tabular}\\
			
			& \begin{tabular}[c]{@{}c@{}}$+$0.5\end{tabular} & \begin{tabular}[c]{@{}c@{}}0.08\end{tabular} & \begin{tabular}[c]{@{}c@{}}0.30\end{tabular} & \begin{tabular}[c]{@{}c@{}}$-$0.96\end{tabular} & \begin{tabular}[c]{@{}c@{}}$-$0.10\end{tabular} & \begin{tabular}[c]{@{}c@{}}$-$0.40\end{tabular}\\
			
			& \begin{tabular}[c]{@{}c@{}}$+$1.0\end{tabular} & \begin{tabular}[c]{@{}c@{}}0.09\end{tabular} & \begin{tabular}[c]{@{}c@{}}0.35\end{tabular} & \begin{tabular}[c]{@{}c@{}}$-$0.98\end{tabular} & \begin{tabular}[c]{@{}c@{}}$-$0.08\end{tabular} & \begin{tabular}[c]{@{}c@{}}$-$0.33\end{tabular}\\
			\colrule
			& \begin{tabular}[c]{@{}c@{}}$-$1.0\end{tabular} & \begin{tabular}[c]{@{}c@{}}$-$0.31\end{tabular} & \begin{tabular}[c]{@{}c@{}}$-$0.03\end{tabular} & \begin{tabular}[c]{@{}c@{}}$-$0.04\end{tabular} & \begin{tabular}[c]{@{}c@{}}0.02\end{tabular} & \begin{tabular}[c]{@{}c@{}}0.16\end{tabular}\\
			
			& \begin{tabular}[c]{@{}c@{}}$-$0.5\end{tabular} & \begin{tabular}[c]{@{}c@{}}$-$0.34\end{tabular} & \begin{tabular}[c]{@{}c@{}}$-$0.02\end{tabular} & \begin{tabular}[c]{@{}c@{}}0.01\end{tabular} & \begin{tabular}[c]{@{}c@{}}0.02\end{tabular} & \begin{tabular}[c]{@{}c@{}}0.27\end{tabular}\\
			
			fcc-Ru/Fe/Ir & \begin{tabular}[c]{@{}c@{}}0.0\end{tabular} & \begin{tabular}[c]{@{}c@{}}$-$0.36\end{tabular} & \begin{tabular}[c]{@{}c@{}}$-$0.01\end{tabular} & \begin{tabular}[c]{@{}c@{}}0.06\end{tabular} & \begin{tabular}[c]{@{}c@{}}0.02\end{tabular} & \begin{tabular}[c]{@{}c@{}}0.37\end{tabular}\\
			
			& \begin{tabular}[c]{@{}c@{}}$+$0.5\end{tabular} & \begin{tabular}[c]{@{}c@{}}$-$0.38\end{tabular} & \begin{tabular}[c]{@{}c@{}}0.01\end{tabular} & \begin{tabular}[c]{@{}c@{}}0.10\end{tabular} & \begin{tabular}[c]{@{}c@{}}0.03\end{tabular} & \begin{tabular}[c]{@{}c@{}}0.45\end{tabular}\\
			
			& \begin{tabular}[c]{@{}c@{}}$+$1.0\end{tabular} & \begin{tabular}[c]{@{}c@{}}$-$0.40\end{tabular} & \begin{tabular}[c]{@{}c@{}}0.03\end{tabular} & \begin{tabular}[c]{@{}c@{}}0.13\end{tabular} & \begin{tabular}[c]{@{}c@{}}0.03\end{tabular} & \begin{tabular}[c]{@{}c@{}}0.50\end{tabular}\\
			
			\colrule
			& \begin{tabular}[c]{@{}c@{}}$-$1.0\end{tabular} & \begin{tabular}[c]{@{}c@{}}$-$0.56\end{tabular} & \begin{tabular}[c]{@{}c@{}}$-$0.26\end{tabular} & \begin{tabular}[c]{@{}c@{}}$-$0.81\end{tabular} & \begin{tabular}[c]{@{}c@{}}$-$0.13\end{tabular} & \begin{tabular}[c]{@{}c@{}}0.27\end{tabular}\\
			
			& \begin{tabular}[c]{@{}c@{}}$-$0.5\end{tabular} & \begin{tabular}[c]{@{}c@{}}$-$0.65\end{tabular} & \begin{tabular}[c]{@{}c@{}}$-$0.27\end{tabular} & \begin{tabular}[c]{@{}c@{}}$-$0.78\end{tabular} & \begin{tabular}[c]{@{}c@{}}$-$0.15\end{tabular} & \begin{tabular}[c]{@{}c@{}}0.31\end{tabular}\\
			
			hcp-Ru/Fe/Ir & \begin{tabular}[c]{@{}c@{}}0.0\end{tabular} & \begin{tabular}[c]{@{}c@{}}$-$0.75\end{tabular} & \begin{tabular}[c]{@{}c@{}}$-$0.28\end{tabular} & \begin{tabular}[c]{@{}c@{}}$-$0.73\end{tabular} & \begin{tabular}[c]{@{}c@{}}$-$0.16\end{tabular} & \begin{tabular}[c]{@{}c@{}}0.34\end{tabular}\\
			
			& \begin{tabular}[c]{@{}c@{}}$+$0.5\end{tabular} & \begin{tabular}[c]{@{}c@{}}$-$0.84\end{tabular} & \begin{tabular}[c]{@{}c@{}}$-$0.30\end{tabular} & \begin{tabular}[c]{@{}c@{}}$-$0.66\end{tabular} & \begin{tabular}[c]{@{}c@{}}$-$0.16\end{tabular} & \begin{tabular}[c]{@{}c@{}}0.36\end{tabular}\\
			
			& \begin{tabular}[c]{@{}c@{}}$+$1.0\end{tabular} & \begin{tabular}[c]{@{}c@{}}$-$0.93\end{tabular} & \begin{tabular}[c]{@{}c@{}}$-$0.32\end{tabular} & \begin{tabular}[c]{@{}c@{}}$-$0.59\end{tabular} & \begin{tabular}[c]{@{}c@{}}$-$0.16\end{tabular} & \begin{tabular}[c]{@{}c@{}}0.37\end{tabular}\\
			
		\end{tabular}
		\label{apptab:table1}
	\end{ruledtabular}
\end{table}

The increase (decrease) of the spin-up charge density at positive (negative) electric field near the Rh and Ir layers (Fig.~\ref{fig:f8}(b)) is caused by the $d_{xz}$ and $d_{z^2}$ orbitals, which are more localized at the atomic sites. The spin-majority LDOS of the Rh 4$d_{xz}$ (Fig.~\ref{fig:f9}(b)) and 4$d_{z^2}$ (Fig.~\ref{fig:f9}(c)) orbitals at around 0.5 eV below the Fermi energy and Ir 5$d_{xz}$ (Fig.~\ref{fig:f9}(h)) and 5$d_{z^2}$ (Fig.~\ref{fig:f9}(i)) orbitals at around 0.5 and 1.3 eV below the Fermi energy, respectively, increases (decreases) at the positive (negative) electric field with reference to the zero field states, leading to an enhancement (reduction) of the spin-up charge density. The spin-up charge of the Rh and Ir $d$ orbitals at positive (negative) electric field increases (decreases) by 13.8 and 6.7$\times 10^{-3}$ electrons, respectively (8.2 and 3.3$\times 10^{-3}$ electrons, respectively) with respect to the zero field value (Table~\ref{tab:table4}), which supports the above field-induced variation of the LDOS.   

The pronounced spin-up and spin-down charge density difference peaks at the Fe layer with opposite sign (Fig.~\ref{fig:f8}(b)) can be attributed to the $d_{x^2-y^2}$, $d_{xz}$, and $d_{z^2}$ orbitals. The spin-majority states of $d_{x^2-y^2}$ orbitals at around 1.6 eV below the Fermi energy (Fig.~\ref{fig:f9}(d)) and those of $d_{z^2}$ orbitals at around 0.5 eV below the Fermi energy (Fig.~\ref{fig:f9}(f)) increases (decreases) at positive (negative) electric field with respect to the zero field states, thus explaining the spin-up charge density peak at $\mathcal{E} = \pm 1.0$~V/\AA. The spin-down charge density peak at positive (negative) electric field can be explained by the shift of spin-minority states of the $d_{x^2-y^2}$ (Fig.~\ref{fig:f9}(d)), $d_{xz}$ (Fig.~\ref{fig:f9}(e)), and $d_{z^2}$ (Fig.~\ref{fig:f9}(f)) orbitals towards higher (lower) energy. The prominent zero-field spin-minority peaks of $d_{x^2-y^2}$ and $d_{z^2}$ at the Fermi energy shift in a similar manner with electric field as those of the Rh and Ir $p_{z}$ peaks (Figs.~\ref{fig:f9}(a) and \ref{fig:f9}(g)), indicating a hybridization of these states. The shift of spin-minority states of $d_{xz}$ character (Fig.~\ref{fig:f9}(e)) is not very prominent at the Fermi energy, but it is clear at around 0.5 eV above the Fermi energy.

The field-induced modification in the integrated charge of the Fe 3$d$ orbitals is consistent with the change in LDOS with the electric field (Table~\ref{tab:table4}). The spin-up and spin-down charge of the Fe 3$d$ orbitals at positive electric field increases by 17.3 $\times 10^{-3}$ electrons and decreases by 22.6$\times 10^{-3}$ electrons, respectively, with respect to the zero field value, while that at the negative electric field decreases by 12.4$\times 10^{-3}$ electrons and increases by 16.9$\times 10^{-3}$ electrons, respectively. Note that the changes in the occupation of the Fe 4$s$ spin-up and spin-down electrons possess the same sign as those of the 3$d$ electrons due to $sd$ hybridization. Moreover, the modified charge of the Fe 3$d$ orbitals is strongly spin-polarized, leading to the spin-polarization of the charge density difference within the trilayer. In contrast, the charges of the 5$p$ orbitals of Rh and 6$p$ orbitals of Ir are weakly spin-polarized resulting in a nearly unpolarized charge density difference in the vacuum (Fig.~\ref{fig:f8}(b)).

Similarly, the charge density difference for the hcp-Pd/Fe/Ir (Fig.~\ref{fig:f8}(a)) and hcp-Ru/Fe/Ir (Fig.~\ref{fig:f8}(c)) trilayers can be explained from field-induced changes in spin-majority and minority LDOS of the constituting atoms (not shown) and electronic occupation of orbitals (Table~\ref{apptab:table3} and Table~\ref{apptab:table4} in Appendix~\ref{appsec:orb_chg}). Specifically, the spin-up and spin-down charge density peak at the Fe layer for hcp-Ru/Fe/Ir (Fig.~\ref{fig:f8}(c)) exhibits opposite behavior compared to those for the Pd/Fe/Ir and Rh/Fe/Ir trilayers (Figs.~\ref{fig:f8}(a) and~\ref{fig:f8}(b)), which is consistent with field-induced changes in the electronic occupation of the Fe 3$d$ orbitals (Table~\ref{apptab:table4}). The spin-up and spin-down charge of the Fe 3$d$ orbitals at positive electric field decreases by 3.7$\times 10^{-3}$ electrons and increases by 2.3$\times 10^{-3}$ electrons, respectively, with respect to the zero field value, while that at negative electric field increases by 5.5$\times 10^{-3}$ electrons and decreases by 4.3$\times 10^{-3}$ electrons, respectively. Furthermore, the reduced height of the Fe peaks for hcp-Ru/Fe/Ir is associated with a smaller field-induced change in Fe 3$d$ occupation compared to the Pd/Fe/Ir and Rh/Fe/Ir trilayers. The variation in the electronic charge of the 3$d$ orbitals for hcp-Ru/Fe/Ir is nearly one order of magnitude smaller than hcp-Rh/Fe/Ir (Table~\ref{tab:table4}) and approximately 2--5 times smaller than hcp-Pd/Fe/Ir (Table~\ref{apptab:table3}).

The field-induced modification in the magnetic interaction constants for the trilayers can be related to the changes of the spin-dependent LDOS under the electric field. The shift of the Fe LDOS under an external electric field indicates modification in the pairwise exchange coupling, as explained in Ref.~\cite{bauer2013}, for an Fe nano-island on Ir(111). This shift also modifies the LDOS at the Fermi energy, which affects the cyclic hopping of electrons associated with the HOI, leading to a change in their coupling strength~\cite{hoffmann20}.

\begin{table} [!htbp]
	\centering
	\caption{\justifying Energy differences ($\Delta E$) for calculation of HOI. Energy difference between the multi-$Q$ ($uudd$ and 3$Q$) and single-$Q$ (spin spiral) states for fcc-Pd/Fe/Ir and hcp-Pd/Fe/Ir trilayers at $\mathcal{E}= 0,\pm 0.3,\pm 0.6$~V/\AA~and fcc-Rh/Fe/Ir, hcp-Rh/Fe/Ir, fcc-Ru/Fe/Ir, and hcp-Ru/Fe/Ir trilayers at $\mathcal{E}= 0, \pm 0.5, \pm 1.0$~V/\AA~used to calculate the HOI constants (Tables~\ref{tab:table1},~\ref{tab:table2}, and \ref{tab:table3}). $\Delta E$s are defined in Eqs.~(8)--(10)) and expressed in meV.}
	\begin{ruledtabular}
		\begin{tabular}{ccccc}
			system & $\mathcal{E}$ & $\Delta E_{3\overline{\mathrm{K}}/4}^{uudd}$ &  $\Delta E_{\overline{\mathrm{M}}/2}^{uudd}$ & $\Delta E_{\overline{\mathrm{M}}}^{3Q}$\\
			\colrule
			& $-$0.6 & \begin{tabular}[c]{@{}c@{}}23.52\end{tabular} & \begin{tabular}[c]{@{}c@{}}14.42\end{tabular} & \begin{tabular}[c]{@{}c@{}}47.66\end{tabular}\\
			
			& $-$0.3 & \begin{tabular}[c]{@{}c@{}}24.57\end{tabular} & \begin{tabular}[c]{@{}c@{}}14.85\end{tabular} & \begin{tabular}[c]{@{}c@{}}48.01\end{tabular}\\
			
			fcc-Pd/Fe/Ir &  0.0 & \begin{tabular}[c]{@{}c@{}}25.55\end{tabular} & \begin{tabular}[c]{@{}c@{}}15.28\end{tabular} & \begin{tabular}[c]{@{}c@{}}48.36\end{tabular}\\
			
			& $+$0.3 & \begin{tabular}[c]{@{}c@{}}26.44\end{tabular} & \begin{tabular}[c]{@{}c@{}}15.65\end{tabular} & \begin{tabular}[c]{@{}c@{}}48.67\end{tabular}\\
			
			& $+$0.6 & \begin{tabular}[c]{@{}c@{}}27.28\end{tabular} & \begin{tabular}[c]{@{}c@{}}16.03\end{tabular} & \begin{tabular}[c]{@{}c@{}}48.97\end{tabular}\\
			\colrule
			& $-$0.6 & \begin{tabular}[c]{@{}c@{}}17.91\end{tabular} & \begin{tabular}[c]{@{}c@{}}$-$3.40\end{tabular} & \begin{tabular}[c]{@{}c@{}}41.21\end{tabular}\\
			
			& $-$0.3 & \begin{tabular}[c]{@{}c@{}}19.19\end{tabular} & \begin{tabular}[c]{@{}c@{}}$-$3.97\end{tabular} & \begin{tabular}[c]{@{}c@{}}42.31\end{tabular}\\
			
			hcp-Pd/Fe/Ir & 0.0 & \begin{tabular}[c]{@{}c@{}}20.41\end{tabular} & \begin{tabular}[c]{@{}c@{}}$-$4.59\end{tabular} & \begin{tabular}[c]{@{}c@{}}43.37\end{tabular}\\
			
			& $+$0.3 & \begin{tabular}[c]{@{}c@{}}21.58\end{tabular} & \begin{tabular}[c]{@{}c@{}}$-$5.28\end{tabular} & \begin{tabular}[c]{@{}c@{}}44.38\end{tabular}\\
			
			& $+$0.6 & \begin{tabular}[c]{@{}c@{}}22.69\end{tabular} & \begin{tabular}[c]{@{}c@{}}$-$6.01\end{tabular} & \begin{tabular}[c]{@{}c@{}}45.34\end{tabular}\\
			\colrule
			& $-$1.0 & \begin{tabular}[c]{@{}c@{}}19.34\end{tabular} & \begin{tabular}[c]{@{}c@{}}13.95\end{tabular} & \begin{tabular}[c]{@{}c@{}}44.80\end{tabular}\\
			
			& $-$0.5 & \begin{tabular}[c]{@{}c@{}}19.57\end{tabular} & \begin{tabular}[c]{@{}c@{}}13.47\end{tabular} & \begin{tabular}[c]{@{}c@{}}44.89\end{tabular}\\
			
			fcc-Rh/Fe/Ir &  0.0 & \begin{tabular}[c]{@{}c@{}}19.67\end{tabular} & \begin{tabular}[c]{@{}c@{}}12.95\end{tabular} & \begin{tabular}[c]{@{}c@{}}44.88\end{tabular}\\
			
			& $+$0.5 & \begin{tabular}[c]{@{}c@{}}19.73\end{tabular} & \begin{tabular}[c]{@{}c@{}}12.46\end{tabular} & \begin{tabular}[c]{@{}c@{}}44.76\end{tabular}\\
			
			& $+$1.0 & \begin{tabular}[c]{@{}c@{}}19.68\end{tabular} & \begin{tabular}[c]{@{}c@{}}11.86\end{tabular} & \begin{tabular}[c]{@{}c@{}}44.56\end{tabular}\\
			\colrule
			& $-$1.0 & \begin{tabular}[c]{@{}c@{}}21.28\end{tabular} & \begin{tabular}[c]{@{}c@{}}$-$17.73\end{tabular} & \begin{tabular}[c]{@{}c@{}}33.62\end{tabular}\\
			
			& $-$0.5 & \begin{tabular}[c]{@{}c@{}}22.14\end{tabular} & \begin{tabular}[c]{@{}c@{}}$-$18.75\end{tabular} & \begin{tabular}[c]{@{}c@{}}33.78\end{tabular}\\
			
			hcp-Rh/Fe/Ir &  0.0 & \begin{tabular}[c]{@{}c@{}}22.86\end{tabular} & \begin{tabular}[c]{@{}c@{}}$-$19.65\end{tabular} & \begin{tabular}[c]{@{}c@{}}33.86\end{tabular}\\
			
			& $+$0.5 & \begin{tabular}[c]{@{}c@{}}23.43\end{tabular} & \begin{tabular}[c]{@{}c@{}}$-$20.41\end{tabular} & \begin{tabular}[c]{@{}c@{}}33.88\end{tabular}\\
			
			& $+$1.0 & \begin{tabular}[c]{@{}c@{}}23.80\end{tabular} & \begin{tabular}[c]{@{}c@{}}$-$21.05\end{tabular} & \begin{tabular}[c]{@{}c@{}}33.84\end{tabular}\\
			\colrule
			& $-$1.0 & \begin{tabular}[c]{@{}c@{}}12.77\end{tabular} & \begin{tabular}[c]{@{}c@{}}2.69\end{tabular} & \begin{tabular}[c]{@{}c@{}}21.97\end{tabular}\\
			
			& $-$0.5 & \begin{tabular}[c]{@{}c@{}}12.85\end{tabular} & \begin{tabular}[c]{@{}c@{}}2.88\end{tabular} & \begin{tabular}[c]{@{}c@{}}22.40\end{tabular}\\
			
			fcc-Ru/Fe/Ir &  0.0 & \begin{tabular}[c]{@{}c@{}}12.63\end{tabular} & \begin{tabular}[c]{@{}c@{}}2.84\end{tabular} & \begin{tabular}[c]{@{}c@{}}22.83\end{tabular}\\
			
			& $+$0.5 & \begin{tabular}[c]{@{}c@{}}12.40\end{tabular} & \begin{tabular}[c]{@{}c@{}}2.89\end{tabular} & \begin{tabular}[c]{@{}c@{}}23.22\end{tabular}\\
			
			& $+$1.0 & \begin{tabular}[c]{@{}c@{}}12.09\end{tabular} & \begin{tabular}[c]{@{}c@{}}2.94\end{tabular} & \begin{tabular}[c]{@{}c@{}}23.60\end{tabular}\\
			\colrule
			& $-$1.0 & \begin{tabular}[c]{@{}c@{}}26.19\end{tabular} & \begin{tabular}[c]{@{}c@{}}$-$2.10\end{tabular} & \begin{tabular}[c]{@{}c@{}}19.71\end{tabular}\\
			
			& $-$0.5 & \begin{tabular}[c]{@{}c@{}}26.10\end{tabular} & \begin{tabular}[c]{@{}c@{}}$-$1.28\end{tabular} & \begin{tabular}[c]{@{}c@{}}21.07\end{tabular}\\
			
			hcp-Ru/Fe/Ir &  0.0 & \begin{tabular}[c]{@{}c@{}}25.67\end{tabular} & \begin{tabular}[c]{@{}c@{}}$-$0.59\end{tabular} & \begin{tabular}[c]{@{}c@{}}22.18\end{tabular}\\
			
			& $+$0.5 & \begin{tabular}[c]{@{}c@{}}24.90\end{tabular} & \begin{tabular}[c]{@{}c@{}}$-$0.09\end{tabular} & \begin{tabular}[c]{@{}c@{}}22.99\end{tabular}\\
			
			& $+$1.0 & \begin{tabular}[c]{@{}c@{}}23.98\end{tabular} & \begin{tabular}[c]{@{}c@{}}$-$0.29\end{tabular} & \begin{tabular}[c]{@{}c@{}}23.51\end{tabular}\\
			
		\end{tabular}
		\label{apptab:table2}
	\end{ruledtabular}
\end{table}

\begin{table*}[!htbp]
\centering
	\caption{\justifying Variation of orbital charge for hcp-Pd/Fe/Ir with electric fields. Difference in electron occupation of Pd 5$s$, 5$p$, and 4$d$, Fe 4$s$, 4$p$, and 3$d$, and Ir 6$s$, 6$p$, and 5$d$ valence orbitals at $\mathcal{E} = \pm 0.6$~V/\AA~with reference to $\mathcal{E} = 0.0$~V/\AA. The electronic occupation is shown for spin-up (majority) and spin-down (minority) channels and it is calculated from the integrated charge within the muffin-tin sphere. The electric field strength $\mathcal{E}$ is expressed in V/\AA.}
    \label{apptab:table3}
\begin{ruledtabular}

\begin{tabular}{cc ccc ccc ccc}

& &
\multicolumn{3}{c}{Pd ($\times10^{-3}$)} &
\multicolumn{3}{c}{Fe ($\times10^{-3}$)} &
\multicolumn{3}{c}{Ir ($\times10^{-3}$)} \\

\cline{3-5} \cline{6-8} \cline{9-11}

$\mathcal{E}$ & spin &
$5s$ & $5p$ & $4d$ &
$4s$ & $4p$ & $3d$ &
$6s$ & $6p$ & $5d$ \\

\midrule

 0.6 & up   
& $-$0.3 & $-$0.7 &  4.4
&  0.2 &  0.5 &  7.3
& $-$0.05 & $-$0.4 &  4.8 \\

 0.6 & down 
& $-$0.5 & $-$0.9 & $-$3.8
& $-$0.2 &  0.7 & $-$10.9
& $-$0.3 & $-$1.1 & $-$4.5 \\

\midrule

$-$0.6 & up   
&  0.2 &  0.6 & $-$4.3
& $-$0.2 & $-$0.5 & $-$7.3
& $-$0.09 &  0.4 & $-$4.0 \\

$-$0.6 & down 
&  0.3 &  0.9 &  3.6
&  0.2 & $-$0.7 & 10.9
&  0.2 &  0.9 &  4.1 \\

\end{tabular}
\end{ruledtabular}
\end{table*}

\section{\label{sec:conc} Conclusion}

In this work, we have systematically studied the magnetic exchange interactions of unsupported 4$d$/Fe/Ir trilayers under electric fields based on DFT. We have selected Pd/Fe/Ir and Rh/Fe/Ir trilayers due to experimental observations of complex noncollinear spin states in the corresponding
ultrathin films at surfaces, and Ru/Fe/Ir trilayers to study the effect of band filling of the $4d$
transition metal. Both hcp and fcc stackings of the 4$d$ overlayer have been considered, since these have been observed experimentally. The Heisenberg pairwise exchange interactions and higher-order exchange interactions, arising in fourth-order perturbation theory, are calculated and analyzed for these transition-metal trilayers for various strengths of the external electric field.

\begin{table*}[!htbp]
\centering
	\caption{\justifying Variation of orbital charge for hcp-Ru/Fe/Ir with electric fields. Difference in electron occupation of Ru 5$s$, 5$p$, and 4$d$, Fe 4$s$, 4$p$, and 3$d$, and Ir 6$s$, 6$p$, and 5$d$ valence orbitals at $\mathcal{E} = \pm 1.0$~V/\AA~with reference to $\mathcal{E} = 0.0$~V/\AA. The electronic occupation is shown for spin-up (majority) and spin-down (minority) channels and it is calculated from the integrated charge within the muffin-tin sphere. The electric field strength $\mathcal{E}$ is expressed in V/\AA.}
    \label{apptab:table4}
\begin{ruledtabular}

\begin{tabular}{cc ccc ccc ccc}
& &
\multicolumn{3}{c}{Ru ($\times10^{-3}$)} &
\multicolumn{3}{c}{Fe ($\times10^{-3}$)} &
\multicolumn{3}{c}{Ir ($\times10^{-3}$)} \\

\cline{3-5} \cline{6-8} \cline{9-11}

$\mathcal{E}$ & spin &
$5s$ & $5p$ & $4d$ &
$4s$ & $4p$ & $3d$ &
$6s$ & $6p$ & $5d$ \\

\midrule

 1.0 & up   
& $-$0.5 & $-$1.0 & $-$3.1
& $-$0.04 &  0.3 & $-$3.7
& $-$0.4 & $-$1.0 & $-$2.7 \\

 1.0 & down 
& $-$0.3 & $-$0.7 &  1.4
& $-$0.1 &  0.5 &  2.3
& $-$0.2 & $-$0.9 & $-$0.2 \\

\midrule

$-$1.0 & up   
&  0.2 &  0.9 &  2.6
&  0.06 & $-$0.1 &  5.5
&  0.1 &  1.0 &  2.8 \\

$-$1.0 & down 
&  0.05 &  0.6 & $-$1.1
&  0.04 & $-$0.5 & $-$4.3
&  0.0 &  0.8 & $-$0.4 \\

\end{tabular}
\end{ruledtabular}
\end{table*}

We find that the energy dispersion of spin spirals and the total energies of multi-$Q$ states vary in a similar way for all trilayers with an applied electric field. While the stability of collinear and noncollinear spin states is modified with electric field, the magnetic ground states of all trilayers remain unchanged.
 
The pairwise exchange constants are calculated by mapping the total DFT energy of spin spirals onto the Heisenberg model and the higher-order exchange constants are obtained from the total energy differences between multi-$Q$ and single-$Q$ states, at zero and finite electric fields for all trilayers. The signs of the interaction constants remain unchanged with electric fields. Their values follow similar trends with electric field for the Pd/Fe/Ir and Rh/Fe/Ir trilayers, however, the trends differ for Ru/Fe/Ir. The relative changes of all interaction constants exhibit a linear variation up to moderate electric field values of about 0.5 V/{\AA}~for all trilayers and deviate at higher values of the electric field for the Rh/Fe/Ir and Ru/Fe/Ir trilayers.

Analysis of the spin-resolved local density of states reveals that the electric-field induced variation of the vacuum charge density in front of the top and bottom layers is caused by charge depletion or accumulation in the $p_{z}$ and $d_{z^2}$ orbitals, while the spin-dependent screening of the electric field near the top and bottom layers and at the Fe layer is governed by $d$ orbitals. These variations in the local density of states result in a modification of the pairwise and higher-order exchange interactions.

Our study shows that exchange interactions can be tuned in magnetic transition-metal trilayers by
applying an electric field. Combining with previous finding that the exchange and higher-order exchange interactions play a crucial role in the stability of topological spin structures, such as skyrmions and antiskyrmions, at transition-metal interfaces, we conclude that the contribution of these interactions can be significant in electric field control and switching of the topological magnetic states at the interfaces. We anticipate that our study will stimulate experiments on the corresponding film systems.

\section{\label{sec:ackn} Acknowledgments}

S.P. graciously acknowledges IISER Thiruvananthapuram for funding and computing time on the Padmanabha cluster. He also acknowledges funding from the Anusandhan National Research Foundation (ANRF/ECRG/2024/001865/PMS). The authors thank Mara Gutzeit for providing the input files for the trilayers, which served as the starting point of the electric field calculations.

\appendix

\section{\label{appsec:beyond_j3} Field-induced variation of the beyond third nearest-neighbor exchange constants}

In the results and discussion section (Sec.~\ref{sec:resdiss}), we studied the effect of an electric field on the first ($J_1$), second ($J_2$), and third ($J_3$) nearest-neighbor exchange interaction constants for three series of trilayers, e.g., Pd/Fe/Ir, Rh/Fe/Ir, and Ru/Fe/Ir with hcp and fcc stackings of the overlayers, and systematically analyze their trends. However, for completeness, we also calculated the pairwise exchange constants up to the eight nearest-neighbor at the considered values of the electric field for all six trilayers, which are listed in Table~\ref{apptab:table1}.

\section{\label{appsec:multiq-singleq} Energy difference between the multi-$Q$ and single-$Q$ states at various electric fields}

In Sec. III, results and discussion, we have analyzed the field-induced trends of the three HOI constants ($B_1$, $Y_1$, and $K_1$) for three series of trilayers, e.g., Pd/Fe/Ir, Rh/Fe/Ir, and Ru/Fe/Ir with hcp and fcc stackings of the overlayers. The energy difference between the single-$Q$ (spin spiral) and multi-$Q$ states ($uudd$ and 3$Q$) for the six trilayers used to calculate those HOI constants is listed in Table~\ref{apptab:table2} at all electric fields considered. The HOI constants are computed using Eqs.~(5)--(10).

\section{\label{appsec:orb_chg} Field-induced variation of orbital charge}

In Sec.~\ref{subsec:spin_screening}, spin-dependent screening of the electric field, we relate changes in the electronic occupations of atomic orbitals of the hcp-Rh/Fe/Ir trilayers under the electric field to the spin-dependent variation of the charge density difference. For completeness, we also listed the field-induced variation of electronic occupation of $s$, $p$, and $d$ orbitals of each constituting atom of the hcp-Pd/Fe/Ir trilayer in Table~\ref{apptab:table3} and those of the hcp-Ru/Fe/Ir trilayer in Table~\ref{apptab:table4}. 
 
%
 
\end{document}